\pdfoutput=1
\documentclass[11pt,twoside,a4paper,pdftex,cmspaper,final,collab]{cms-tdr}

\begin{document}\cmsNoteHeader{CFT-09-010}
%
%
%

%
%
\hyphenation{env-iron-men-tal}
\hyphenation{had-ron-i-za-tion}
\hyphenation{cal-or-i-me-ter}
\hyphenation{de-vices}
%
%
\RCS$Revision: 1.23 $
\RCS$Date: 2009/12/29 09:34:31 $
\RCS$Name:  $
%
%
%

\newcommand {\etal}{\mbox{et al.}\xspace} 
\newcommand {\ie}{\mbox{i.e.}\xspace}     
\newcommand {\eg}{\mbox{e.g.}\xspace}     
\newcommand {\etc}{\mbox{etc.}\xspace}     
\newcommand {\vs}{\mbox{\sl vs.}\xspace}      
\newcommand {\mdash}{\ensuremath{\mathrm{-}}} 

\newcommand {\Lone}{Level-1\xspace} 
\newcommand {\Ltwo}{Level-2\xspace}
\newcommand {\Lthree}{Level-3\xspace}

\providecommand{\ACERMC} {\textsc{AcerMC}\xspace}
\providecommand{\ALPGEN} {{\textsc{alpgen}}\xspace}
\providecommand{\CHARYBDIS} {{\textsc{charybdis}}\xspace}
\providecommand{\CMKIN} {\textsc{cmkin}\xspace}
\providecommand{\CMSIM} {{\textsc{cmsim}}\xspace}
\providecommand{\CMSSW} {{\textsc{cmssw}}\xspace}
\providecommand{\COBRA} {{\textsc{cobra}}\xspace}
\providecommand{\COCOA} {{\textsc{cocoa}}\xspace}
\providecommand{\COMPHEP} {\textsc{CompHEP}\xspace}
\providecommand{\EVTGEN} {{\textsc{evtgen}}\xspace}
\providecommand{\FAMOS} {{\textsc{famos}}\xspace}
\providecommand{\GARCON} {\textsc{garcon}\xspace}
\providecommand{\GARFIELD} {{\textsc{garfield}}\xspace}
\providecommand{\GEANE} {{\textsc{geane}}\xspace}
\providecommand{\GEANTfour} {{\textsc{geant4}}\xspace}
\providecommand{\GEANTthree} {{\textsc{geant3}}\xspace}
\providecommand{\GEANT} {{\textsc{geant}}\xspace}
\providecommand{\HDECAY} {\textsc{hdecay}\xspace}
\providecommand{\HERWIG} {{\textsc{herwig}}\xspace}
\providecommand{\HIGLU} {{\textsc{higlu}}\xspace}
\providecommand{\HIJING} {{\textsc{hijing}}\xspace}
\providecommand{\IGUANA} {\textsc{iguana}\xspace}
\providecommand{\ISAJET} {{\textsc{isajet}}\xspace}
\providecommand{\ISAPYTHIA} {{\textsc{isapythia}}\xspace}
\providecommand{\ISASUGRA} {{\textsc{isasugra}}\xspace}
\providecommand{\ISASUSY} {{\textsc{isasusy}}\xspace}
\providecommand{\ISAWIG} {{\textsc{isawig}}\xspace}
\providecommand{\MADGRAPH} {\textsc{MadGraph}\xspace}
\providecommand{\MCATNLO} {\textsc{mc@nlo}\xspace}
\providecommand{\MCFM} {\textsc{mcfm}\xspace}
\providecommand{\MILLEPEDE} {{\textsc{millepede}}\xspace}
\providecommand{\ORCA} {{\textsc{orca}}\xspace}
\providecommand{\OSCAR} {{\textsc{oscar}}\xspace}
\providecommand{\PHOTOS} {\textsc{photos}\xspace}
\providecommand{\PROSPINO} {\textsc{prospino}\xspace}
\providecommand{\PYTHIA} {{\textsc{pythia}}\xspace}
\providecommand{\SHERPA} {{\textsc{sherpa}}\xspace}
\providecommand{\TAUOLA} {\textsc{tauola}\xspace}
\providecommand{\TOPREX} {\textsc{TopReX}\xspace}
\providecommand{\XDAQ} {{\textsc{xdaq}}\xspace}

\newcommand {\DZERO}{D\O\xspace}     


\newcommand{\de}{\ensuremath{^\circ}}
\newcommand{\ten}[1]{\ensuremath{\times \text{10}^\text{#1}}}
\newcommand{\unit}[1]{\ensuremath{\text{\,#1}}\xspace}
\newcommand{\mum}{\ensuremath{\,\mu\text{m}}\xspace}
\newcommand{\micron}{\ensuremath{\,\mu\text{m}}\xspace}
\newcommand{\cm}{\ensuremath{\,\text{cm}}\xspace}
\newcommand{\mm}{\ensuremath{\,\text{mm}}\xspace}
\newcommand{\mus}{\ensuremath{\,\mu\text{s}}\xspace}
\newcommand{\keV}{\ensuremath{\,\text{ke\hspace{-.08em}V}}\xspace}
\newcommand{\MeV}{\ensuremath{\,\text{Me\hspace{-.08em}V}}\xspace}
\newcommand{\GeV}{\ensuremath{\,\text{Ge\hspace{-.08em}V}}\xspace}
\newcommand{\TeV}{\ensuremath{\,\text{Te\hspace{-.08em}V}}\xspace}
\newcommand{\PeV}{\ensuremath{\,\text{Pe\hspace{-.08em}V}}\xspace}
\newcommand{\keVc}{\ensuremath{{\,\text{ke\hspace{-.08em}V\hspace{-0.16em}/\hspace{-0.08em}c}}}\xspace}
\newcommand{\MeVc}{\ensuremath{{\,\text{Me\hspace{-.08em}V\hspace{-0.16em}/\hspace{-0.08em}c}}}\xspace}
\newcommand{\GeVc}{\ensuremath{{\,\text{Ge\hspace{-.08em}V\hspace{-0.16em}/\hspace{-0.08em}c}}}\xspace}
\newcommand{\TeVc}{\ensuremath{{\,\text{Te\hspace{-.08em}V\hspace{-0.16em}/\hspace{-0.08em}c}}}\xspace}
\newcommand{\keVcc}{\ensuremath{{\,\text{ke\hspace{-.08em}V\hspace{-0.16em}/\hspace{-0.08em}c}^\text{2}}}\xspace}
\newcommand{\MeVcc}{\ensuremath{{\,\text{Me\hspace{-.08em}V\hspace{-0.16em}/\hspace{-0.08em}c}^\text{2}}}\xspace}
\newcommand{\GeVcc}{\ensuremath{{\,\text{Ge\hspace{-.08em}V\hspace{-0.16em}/\hspace{-0.08em}c}^\text{2}}}\xspace}
\newcommand{\TeVcc}{\ensuremath{{\,\text{Te\hspace{-.08em}V\hspace{-0.16em}/\hspace{-0.08em}c}^\text{2}}}\xspace}

\newcommand{\pbinv} {\mbox{\ensuremath{\,\text{pb}^\text{$-$1}}}\xspace}
\newcommand{\fbinv} {\mbox{\ensuremath{\,\text{fb}^\text{$-$1}}}\xspace}
\newcommand{\nbinv} {\mbox{\ensuremath{\,\text{nb}^\text{$-$1}}}\xspace}
\newcommand{\percms}{\ensuremath{\,\text{cm}^\text{$-$2}\,\text{s}^\text{$-$1}}\xspace}
\newcommand{\lumi}{\ensuremath{\mathcal{L}}\xspace}
\newcommand{\Lumi}{\ensuremath{\mathcal{L}}\xspace}
%
\newcommand{\LvLow}  {\ensuremath{\mathcal{L}=\text{10}^\text{32}\,\text{cm}^\text{$-$2}\,\text{s}^\text{$-$1}}\xspace}
\newcommand{\LLow}   {\ensuremath{\mathcal{L}=\text{10}^\text{33}\,\text{cm}^\text{$-$2}\,\text{s}^\text{$-$1}}\xspace}
\newcommand{\lowlumi}{\ensuremath{\mathcal{L}=\text{2}\times \text{10}^\text{33}\,\text{cm}^\text{$-$2}\,\text{s}^\text{$-$1}}\xspace}
\newcommand{\LMed}   {\ensuremath{\mathcal{L}=\text{2}\times \text{10}^\text{33}\,\text{cm}^\text{$-$2}\,\text{s}^\text{$-$1}}\xspace}
\newcommand{\LHigh}  {\ensuremath{\mathcal{L}=\text{10}^\text{34}\,\text{cm}^\text{$-$2}\,\text{s}^\text{$-$1}}\xspace}
\newcommand{\hilumi} {\ensuremath{\mathcal{L}=\text{10}^\text{34}\,\text{cm}^\text{$-$2}\,\text{s}^\text{$-$1}}\xspace}


\newcommand{\zp}{\ensuremath{\mathrm{Z}^\prime}\xspace}


\newcommand{\kt}{\ensuremath{k_{\mathrm{T}}}\xspace}
\newcommand{\BC}{\ensuremath{{B_{\mathrm{c}}}}\xspace}
\newcommand{\bbarc}{\ensuremath{{\overline{b}c}}\xspace}
\newcommand{\bbbar}{\ensuremath{{b\overline{b}}}\xspace}
\newcommand{\ccbar}{\ensuremath{{c\overline{c}}}\xspace}
\newcommand{\JPsi}{\ensuremath{{J}/\psi}\xspace}
\newcommand{\bspsiphi}{\ensuremath{B_s \to \JPsi\, \phi}\xspace}
\newcommand{\AFB}{\ensuremath{A_\mathrm{FB}}\xspace}
\newcommand{\EE}{\ensuremath{e^+e^-}\xspace}
\newcommand{\MM}{\ensuremath{\mu^+\mu^-}\xspace}
\newcommand{\TT}{\ensuremath{\tau^+\tau^-}\xspace}
\newcommand{\wangle}{\ensuremath{\sin^{2}\theta_{\mathrm{eff}}^\mathrm{lept}(M^2_\mathrm{Z})}\xspace}
\newcommand{\ttbar}{\ensuremath{{t\overline{t}}}\xspace}
\newcommand{\stat}{\ensuremath{\,\text{(stat.)}}\xspace}
\newcommand{\syst}{\ensuremath{\,\text{(syst.)}}\xspace}

\newcommand{\HGG}{\ensuremath{\mathrm{H}\to\gamma\gamma}}
\newcommand{\gev}{\GeV}
\newcommand{\GAMJET}{\ensuremath{\gamma + \mathrm{jet}}}
\newcommand{\PPTOJETS}{\ensuremath{\mathrm{pp}\to\mathrm{jets}}}
\newcommand{\PPTOGG}{\ensuremath{\mathrm{pp}\to\gamma\gamma}}
\newcommand{\PPTOGAMJET}{\ensuremath{\mathrm{pp}\to\gamma +
\mathrm{jet}
}}
\newcommand{\MH}{\ensuremath{\mathrm{M_{\mathrm{H}}}}}
\newcommand{\RNINE}{\ensuremath{\mathrm{R}_\mathrm{9}}}
\newcommand{\DR}{\ensuremath{\Delta\mathrm{R}}}


\newcommand{\PT}{\ensuremath{p_{\mathrm{T}}}\xspace}
\newcommand{\pt}{\ensuremath{p_{\mathrm{T}}}\xspace}
\newcommand{\ET}{\ensuremath{E_{\mathrm{T}}}\xspace}
\newcommand{\HT}{\ensuremath{H_{\mathrm{T}}}\xspace}
\newcommand{\et}{\ensuremath{E_{\mathrm{T}}}\xspace}
\newcommand{\Em}{\ensuremath{E\!\!\!/}\xspace}
\newcommand{\Pm}{\ensuremath{p\!\!\!/}\xspace}
\newcommand{\PTm}{\ensuremath{{p\!\!\!/}_{\mathrm{T}}}\xspace}
\newcommand{\ETm}{\ensuremath{E_{\mathrm{T}}^{\mathrm{miss}}}\xspace}
\newcommand{\MET}{\ensuremath{E_{\mathrm{T}}^{\mathrm{miss}}}\xspace}
\newcommand{\ETmiss}{\ensuremath{E_{\mathrm{T}}^{\mathrm{miss}}}\xspace}
\newcommand{\VEtmiss}{\ensuremath{{\vec E}_{\mathrm{T}}^{\mathrm{miss}}}\xspace}

%

\newcommand{\ga}{\ensuremath{\gtrsim}}
\newcommand{\la}{\ensuremath{\lesssim}}
\newcommand{\swsq}{\ensuremath{\sin^2\theta_W}\xspace}
\newcommand{\cwsq}{\ensuremath{\cos^2\theta_W}\xspace}
\newcommand{\tanb}{\ensuremath{\tan\beta}\xspace}
\newcommand{\tanbsq}{\ensuremath{\tan^{2}\beta}\xspace}
\newcommand{\sidb}{\ensuremath{\sin 2\beta}\xspace}
\newcommand{\alpS}{\ensuremath{\alpha_S}\xspace}
\newcommand{\alpt}{\ensuremath{\tilde{\alpha}}\xspace}

\newcommand{\QL}{\ensuremath{Q_L}\xspace}
\newcommand{\sQ}{\ensuremath{\tilde{Q}}\xspace}
\newcommand{\sQL}{\ensuremath{\tilde{Q}_L}\xspace}
\newcommand{\ULC}{\ensuremath{U_L^C}\xspace}
\newcommand{\sUC}{\ensuremath{\tilde{U}^C}\xspace}
\newcommand{\sULC}{\ensuremath{\tilde{U}_L^C}\xspace}
\newcommand{\DLC}{\ensuremath{D_L^C}\xspace}
\newcommand{\sDC}{\ensuremath{\tilde{D}^C}\xspace}
\newcommand{\sDLC}{\ensuremath{\tilde{D}_L^C}\xspace}
\newcommand{\LL}{\ensuremath{L_L}\xspace}
\newcommand{\sL}{\ensuremath{\tilde{L}}\xspace}
\newcommand{\sLL}{\ensuremath{\tilde{L}_L}\xspace}
\newcommand{\ELC}{\ensuremath{E_L^C}\xspace}
\newcommand{\sEC}{\ensuremath{\tilde{E}^C}\xspace}
\newcommand{\sELC}{\ensuremath{\tilde{E}_L^C}\xspace}
\newcommand{\sEL}{\ensuremath{\tilde{E}_L}\xspace}
\newcommand{\sER}{\ensuremath{\tilde{E}_R}\xspace}
\newcommand{\sFer}{\ensuremath{\tilde{f}}\xspace}
\newcommand{\sQua}{\ensuremath{\tilde{q}}\xspace}
\newcommand{\sUp}{\ensuremath{\tilde{u}}\xspace}
\newcommand{\suL}{\ensuremath{\tilde{u}_L}\xspace}
\newcommand{\suR}{\ensuremath{\tilde{u}_R}\xspace}
\newcommand{\sDw}{\ensuremath{\tilde{d}}\xspace}
\newcommand{\sdL}{\ensuremath{\tilde{d}_L}\xspace}
\newcommand{\sdR}{\ensuremath{\tilde{d}_R}\xspace}
\newcommand{\sTop}{\ensuremath{\tilde{t}}\xspace}
\newcommand{\stL}{\ensuremath{\tilde{t}_L}\xspace}
\newcommand{\stR}{\ensuremath{\tilde{t}_R}\xspace}
\newcommand{\stone}{\ensuremath{\tilde{t}_1}\xspace}
\newcommand{\sttwo}{\ensuremath{\tilde{t}_2}\xspace}
\newcommand{\sBot}{\ensuremath{\tilde{b}}\xspace}
\newcommand{\sbL}{\ensuremath{\tilde{b}_L}\xspace}
\newcommand{\sbR}{\ensuremath{\tilde{b}_R}\xspace}
\newcommand{\sbone}{\ensuremath{\tilde{b}_1}\xspace}
\newcommand{\sbtwo}{\ensuremath{\tilde{b}_2}\xspace}
\newcommand{\sLep}{\ensuremath{\tilde{l}}\xspace}
\newcommand{\sLepC}{\ensuremath{\tilde{l}^C}\xspace}
\newcommand{\sEl}{\ensuremath{\tilde{e}}\xspace}
\newcommand{\sElC}{\ensuremath{\tilde{e}^C}\xspace}
\newcommand{\seL}{\ensuremath{\tilde{e}_L}\xspace}
\newcommand{\seR}{\ensuremath{\tilde{e}_R}\xspace}
\newcommand{\snL}{\ensuremath{\tilde{\nu}_L}\xspace}
\newcommand{\sMu}{\ensuremath{\tilde{\mu}}\xspace}
\newcommand{\sNu}{\ensuremath{\tilde{\nu}}\xspace}
\newcommand{\sTau}{\ensuremath{\tilde{\tau}}\xspace}
\newcommand{\Glu}{\ensuremath{g}\xspace}
\newcommand{\sGlu}{\ensuremath{\tilde{g}}\xspace}
\newcommand{\Wpm}{\ensuremath{W^{\pm}}\xspace}
\newcommand{\sWpm}{\ensuremath{\tilde{W}^{\pm}}\xspace}
\newcommand{\Wz}{\ensuremath{W^{0}}\xspace}
\newcommand{\sWz}{\ensuremath{\tilde{W}^{0}}\xspace}
\newcommand{\sWino}{\ensuremath{\tilde{W}}\xspace}
\newcommand{\Bz}{\ensuremath{B^{0}}\xspace}
\newcommand{\sBz}{\ensuremath{\tilde{B}^{0}}\xspace}
\newcommand{\sBino}{\ensuremath{\tilde{B}}\xspace}
\newcommand{\Zz}{\ensuremath{Z^{0}}\xspace}
\newcommand{\sZino}{\ensuremath{\tilde{Z}^{0}}\xspace}
\newcommand{\sGam}{\ensuremath{\tilde{\gamma}}\xspace}
\newcommand{\chiz}{\ensuremath{\tilde{\chi}^{0}}\xspace}
\newcommand{\chip}{\ensuremath{\tilde{\chi}^{+}}\xspace}
\newcommand{\chim}{\ensuremath{\tilde{\chi}^{-}}\xspace}
\newcommand{\chipm}{\ensuremath{\tilde{\chi}^{\pm}}\xspace}
\newcommand{\Hone}{\ensuremath{H_{d}}\xspace}
\newcommand{\sHone}{\ensuremath{\tilde{H}_{d}}\xspace}
\newcommand{\Htwo}{\ensuremath{H_{u}}\xspace}
\newcommand{\sHtwo}{\ensuremath{\tilde{H}_{u}}\xspace}
\newcommand{\sHig}{\ensuremath{\tilde{H}}\xspace}
\newcommand{\sHa}{\ensuremath{\tilde{H}_{a}}\xspace}
\newcommand{\sHb}{\ensuremath{\tilde{H}_{b}}\xspace}
\newcommand{\sHpm}{\ensuremath{\tilde{H}^{\pm}}\xspace}
\newcommand{\hz}{\ensuremath{h^{0}}\xspace}
\newcommand{\Hz}{\ensuremath{H^{0}}\xspace}
\newcommand{\Az}{\ensuremath{A^{0}}\xspace}
\newcommand{\Hpm}{\ensuremath{H^{\pm}}\xspace}
\newcommand{\sGra}{\ensuremath{\tilde{G}}\xspace}
\newcommand{\mtil}{\ensuremath{\tilde{m}}\xspace}
\newcommand{\rpv}{\ensuremath{\rlap{\kern.2em/}R}\xspace}
\newcommand{\LLE}{\ensuremath{LL\bar{E}}\xspace}
\newcommand{\LQD}{\ensuremath{LQ\bar{D}}\xspace}
\newcommand{\UDD}{\ensuremath{\overline{UDD}}\xspace}
\newcommand{\Lam}{\ensuremath{\lambda}\xspace}
\newcommand{\Lamp}{\ensuremath{\lambda'}\xspace}
\newcommand{\Lampp}{\ensuremath{\lambda''}\xspace}
\newcommand{\spinbd}[2]{\ensuremath{\bar{#1}_{\dot{#2}}}\xspace}

\newcommand{\MD}{\ensuremath{{M_\mathrm{D}}}\xspace}
\newcommand{\Mpl}{\ensuremath{{M_\mathrm{Pl}}}\xspace}
\newcommand{\Rinv} {\ensuremath{{R}^{-1}}\xspace}

%
%
\hyphenation{en-viron-men-tal}

\cmsNoteHeader{09-010}
\title{Performance Study of the CMS Barrel Resistive Plate Chambers with Cosmic Rays}

\author{the CMS Collaboration}

\date{\today}

\abstract{ In October and November 2008, the CMS collaboration
  conducted a programme of cosmic ray data taking, which has recorded
  about 270 million events. The Resistive Plate Chamber system, which
  is part of the CMS muon detection system, was successfully operated
  in the full barrel. More than 98\,\% of the channels
were operational during the exercise with typical detection efficiency of 90\,\%.
In this paper, the performance of the detector
  during these dedicated runs is reported.  }

\hypersetup{%
pdfauthor={The CMS Collaboration},%
pdftitle={Performance Study of the CMS Barrel Resistive Plate Chambers with Cosmic Rays},%
pdfsubject={CMS},%
pdfkeywords={CMS, RPC, CRAFT}}

\maketitle 


\section{Introduction}

The primary goal of the Compact Muon Solenoid (CMS)
experiment~\cite{cms} is to explore particle physics at the TeV energy scale
exploiting the collisions between protons delivered by the Large
Hadron Collider (LHC)~\cite{lhc}.  The Resistive Plate Chamber (RPC)
system~\cite{rpc} is part of the muon detection system and is used for
triggering purposes.  RPCs have been chosen because of their good time
resolution (about 2~ns) and high granularity, which permit a fast and
efficient triggering of muons over large areas.

During October-November 2008, the CMS Collaboration conducted a month-long data-taking
exercise known as the Cosmic Run At Four Tesla (CRAFT), with the goal
of commissioning
the experiment for extended operation~\cite{CRAFT}. With all installed detector
systems participating,
CMS recorded 270 million cosmic-ray-triggered events with the solenoid
at a central
flux density of 3.8 T.

The muon system is composed of a central barrel (in the
pseudo-rapidity window $|\eta|<1.04$) and two closing endcaps ($1.04
<|\eta|< 2.4$).  The RPCs participated in the CRAFT
data taking with the full barrel and half of the endcaps in
operation. Part of the endcaps was not operational because the readout
electronics was not yet available.

For the RPC system, the CRAFT 2008 exercise has permitted the check of the
full data taking chain, from the detector up to the data acquisition
boards, and the confirmation of the detector performance established during the
quality control phase~\cite{siena}, also in presence of the CMS magnetic
field.
The exercise was also a useful benchmark to complete the commissioning
of the endcap part of the RPC system.

This paper is focused on the barrel RPC system performance.
Sections~\ref{sec:layout} and~\ref{sec:operations} describe the RPC system
layout, and the operation and monitoring procedures during the CRAFT
exercise, respectively. In Section~\ref{sec:synch}, the readout
electronics is presented and the synchronization procedure is
described. Section~\ref{sec:performances} deals with the barrel RPC
performance, in terms of cluster size, position resolution and
detection efficiency.

\section{RPC system layout}\label{sec:layout}

The central feature of the CMS apparatus is a superconducting solenoid
of 6~m internal diameter.  The silicon pixel and strip tracker, the
crystal electromagnetic calorimeter (ECAL) and the brass-scintillator
hadronic calorimeter (HCAL) are located within the solenoidal field
volume.  Muons are measured in gas ionization detectors embedded in
the steel return yoke.
CMS uses a right-handed coordinate system, with the
  origin at the nominal collision point, the $x$-axis pointing to the
  centre of the LHC, the $y$-axis pointing up (perpendicular to the
  LHC plane), and the $z$-axis along the anticlockwise-beam direction.
  The polar angle, $\theta$, is measured from the positive $z$-axis,
  the azimuthal angle, $\phi$, is measured in the $x$-$y$ plane.

The RPC detectors are employed in CMS as a dedicated trigger system in
both the barrel and in the endcap regions.  They complement the muon
tracking system: drift tubes (DT)~\cite{dt} in the barrel and
cathode strip chambers (CSC)~\cite{csc} in the endcaps.

\begin{figure}[hbtp]
  \begin{center}
    \includegraphics[width=0.9\textwidth]{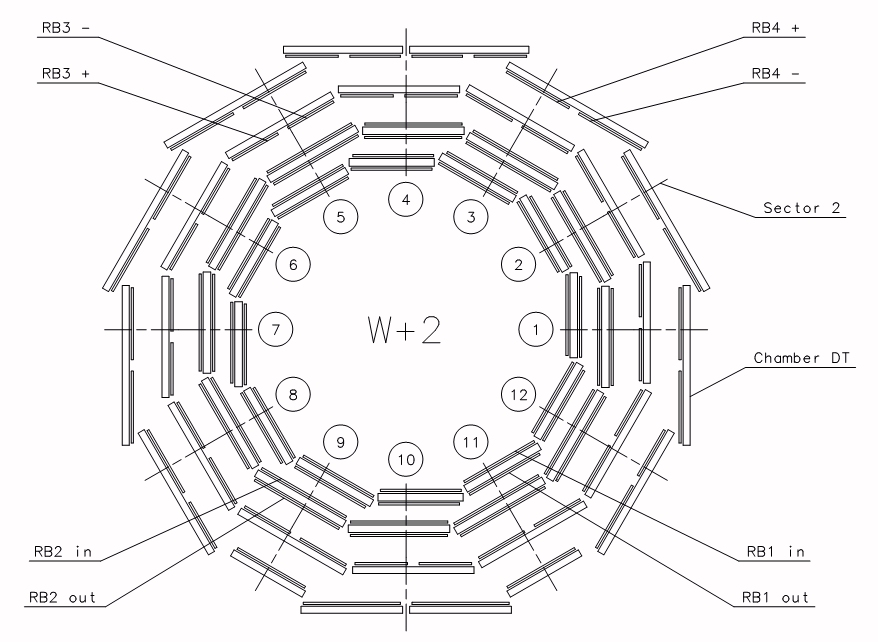}
    \hspace{1cm}
    \caption{Transverse view of the muon system layout in the barrel
      region, showing the positions of the DT and RPC stations.}
    \label{fig:barrelLayout}
  \end{center}
\end{figure}

From the geometrical point of view, the muon system is divided into
five wheels in the barrel and three disks in each endcap.
Each barrel wheel is divided in 12 sectors,
covering the full azimuthal dimension (see Fig.~\ref{fig:barrelLayout}).
Each sector consists of four layers of DTs and six layers of RPCs,
with a total of 480 RPC stations of average area of 12~${\rm m^2}$.
The two innermost DT layers are sandwiched between RPC layers
(RB1$_{\rm in}$ and RB1$_{\rm out}$ for the
innermost DT layer, RB2$_{\rm in}$ and RB2$_{\rm out}$ for the second one).
The third and fourth DT layers are complemented with a single RPC
layer, placed on their inner side (RB3 and RB4).  A detailed description
of the muon system geometry can be found in Ref.~\cite{cms}.

\begin{figure}[hbt]
  \begin{center}
    \includegraphics*[height=7cm, width=0.9\textwidth]{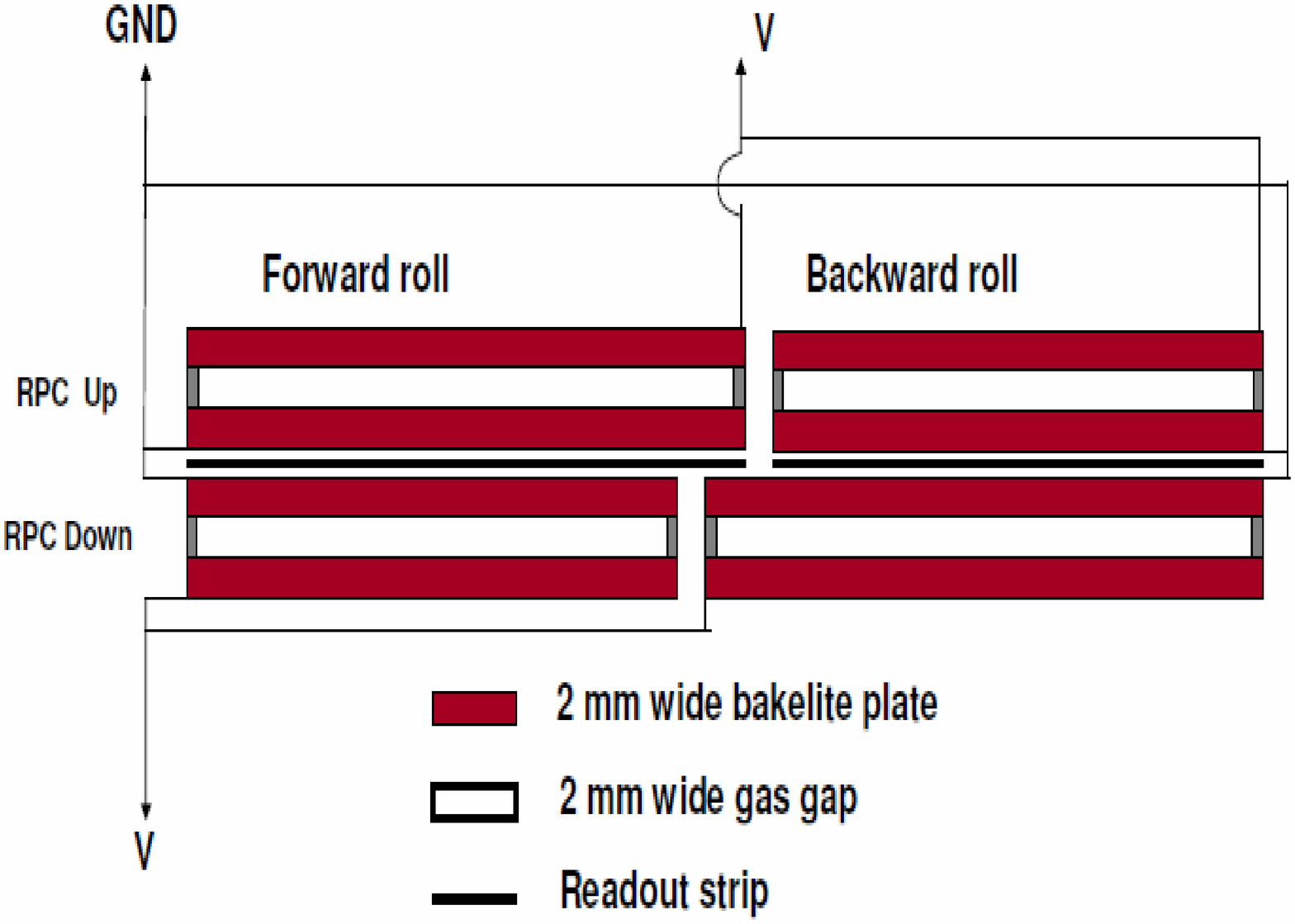}
    \hspace{1cm}
    \caption{Schematic layout of a RPC station composed of two rolls.}
    \label{fig:roll}
  \end{center}
\end{figure}

Each barrel RPC station is mechanically composed of two or three
adjacent RPC units, called ``rolls'', making a total of 1020 rolls for
the barrel RPC system. Figure~\ref{fig:roll} shows a schematic layout of
a RPC station composed by two rolls.  The CMS RPC rolls are double
RPC units, where each unit is composed of two bakelite electrodes,
each 2~mm thick, and a 2~mm wide gas gap.
Readout strips, which are oriented along the beam direction, are
placed between the two RPC units, with an independent electronic
channel for each strip. The total readout system of the barrel RPC
detector consists of 68\,136 channels.

The strip location defines a point in the local reference frame of the RPC
with a precision given by the strip width,
ranging from 2.3~cm, for the innermost RB1$_{\rm in}$ layer, to 4.1~cm, for
the RB4 layer (see Table~\ref{tab:tabclsByLayer}).  The strip signals
are discriminated and shaped into 100~ns digital pulses by
front-end boards with programmable discriminator thresholds.

The RPC detectors, which work in saturated avalanche mode, use a
three-component, non-flammable gas mixture composed of 96.2\,\%
C$_2$H$_2$F$_4$ (R134a), 3.5\,\% isobutane (C$_4$H$_{10}$) and 0.3\,\%
SF$_6$.  Water vapour is added to obtain a mixture with a relative
humidity of 40--50\,\%, in order to maintain a constant bakelite
resistivity and, thus, avoid a degradation of RPC performance under high
background conditions.  The large volume of the total RPC system
(18~m$^3$) and the use of a rather expensive gas mixture impose the
use of a gas recirculation system.  This system consists of the
following modules: the primary supply, the mixer, the humidifier, the
closed-loop circulators, the gas distributors to the chambers, the
purifiers, and the pump. The purifier system is made of a set of
filters designed to remove contaminants from the gas mixture.  The
system operates with a fraction of fresh gas mixture in the range from
10\,\% to 2\,\%.  This is the first time that such a large RPC system is
operated with a closed loop gas system.

\section {Detector operation and monitoring}\label{sec:operations}

The effective working voltage, $V_{\rm eff}$, relevant for the charge
avalanche production inside the RPC, depends on environmental
parameters such as gas temperature (T) and pressure (P), according to
Eq.~\ref{eq:effectiveHV}~\cite{rpc2}

\begin{equation}
\label{eq:effectiveHV}
V_{\rm eff} = V \times \cfrac{P_{0}}{P} \times \cfrac{T}{T_{0}} \quad ,
\end{equation}

where $V_{\rm eff}$ is the effective voltage, $V$ is the applied power
voltage, and $P_{0} = 1010$~mbar and $T_{0} = 293$~K are the reference pressure and
temperature, respectively.  Only the
effective voltage is relevant when comparing detector performance in
different sites and run conditions.

The CMS RPCs have been extensively tested at the production sites and
their physics performance was studied~\cite{rpc,siena}.  During 2006,
a small fraction of the RPC system was calibrated and operated
\cite{MTCC}.  In both cases, it was found that 95\,\% of the maximum
plateau efficiency is reached at an effective voltage of 9.6~kV,
corresponding to an average applied voltage of 9.2~kV for the pressure
conditions in the CMS cavern.

For the CRAFT 2008 exercise, the operating voltage was set to 9.2~kV
and the electronic thresholds of the readout system were set to
230~mV, corresponding to an induced charge of 180~fC.  These operating
conditions are conservative and do not permit the detector to reach
its maximum efficiency. This approach was chosen to maintain low noise
levels and safe operating conditions, since this
was the first time that the full system was operated.  Following
extensive past studies~\cite{rpcTP}, the working conditions were
maintained within strict ranges: temperature lower than
24~${^\circ}\rm{C}$, humidity in the range 40--50\,\% and fraction of
$\rm {O_{2}}$ below 300~ppm.

These parameters were monitored and controlled by the Detector Control
System (DCS)~\cite{dcs}. Their values were stored in a database and
used, offline, to study the system stability.

\subsection {Temperature}

\begin{figure}[hbtp]
\begin{center}
\begin{tabular}{cc}
    \includegraphics[width=0.45\textwidth]{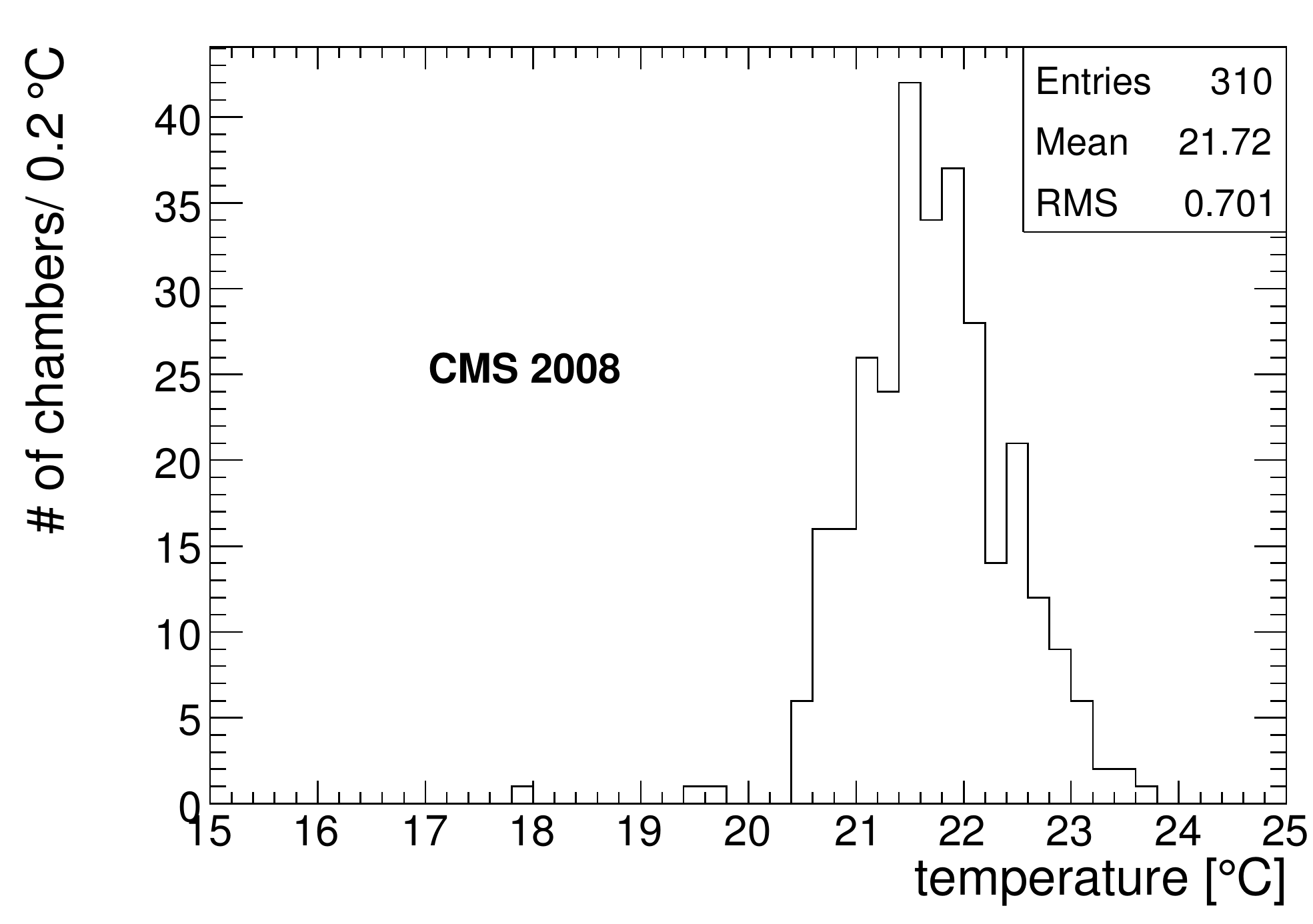} &
    \includegraphics[width=0.45\textwidth]{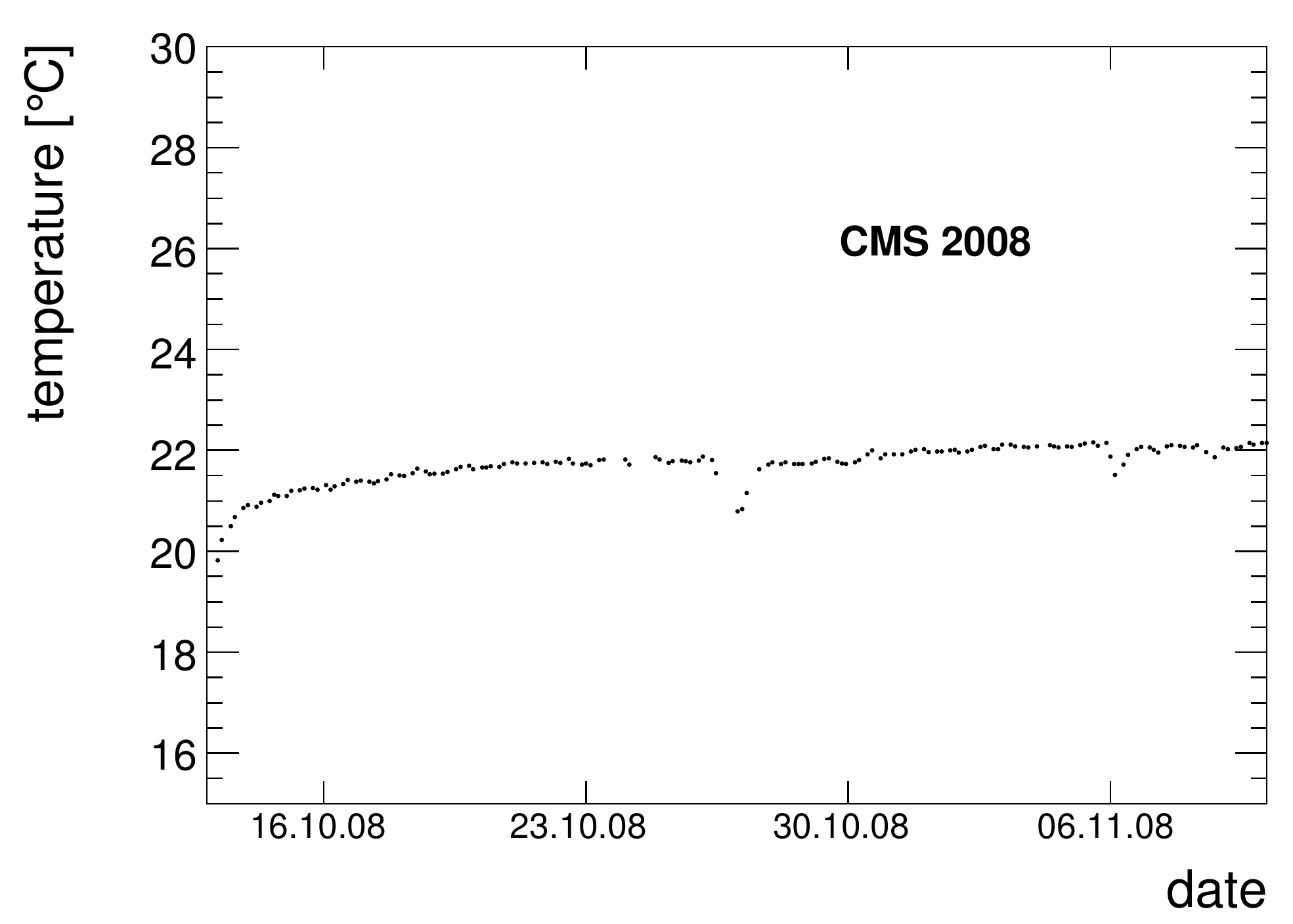} \\
\end{tabular}
  \caption{RPC temperature, as measured by the
      probes installed inside the stations. Left side: distribution of RPC temperature taken at
the end of the CRAFT period and averaged over one full day of data taking.
Right side: Average RPC temperature as a function of time. Each point represents the temperature averaged over 4 hours and over all
      barrel stations. The increase and the variations of the
      temperature are triggered by the start of the CMS
      electronics.}
\label{fig:temperature}
\end{center}
\end{figure}

Out of the 480 RPC stations, 310 are equipped with a temperature probe
installed inside the mechanical frame.  Figure~\ref{fig:temperature}
shows in the left plot the temperature distribution of
those barrel stations at the end of the CRAFT period (averaged
over one day) and on the right plot its average vs. time.  The system
temperature was maintained below 24~${^\circ}{\rm C}$ during the full
CRAFT period, in order to guarantee proper RPC operation.  Clear
variations are observed when the electronics of CMS is switched on or
off. The steep increase seen at the start of the CRAFT exercise,
when all the detectors were fully operating, is a consequence of the
non optimized cooling of the system.  A special effort was dedicated to
increase the cooling circuit capability after the end of the CRAFT
exercise. The temperature of the input cooling water has been reduced by
one degree in order to guarantee a constant stable temperature lower
than 24~${^\circ}$C, as confirmed in 2009, during long
cosmic rays data taking periods.

\subsection {Currents}

\begin{figure}[hbtp]
\begin{center}
\begin{tabular}{cc}
    \includegraphics[width=0.45\textwidth]{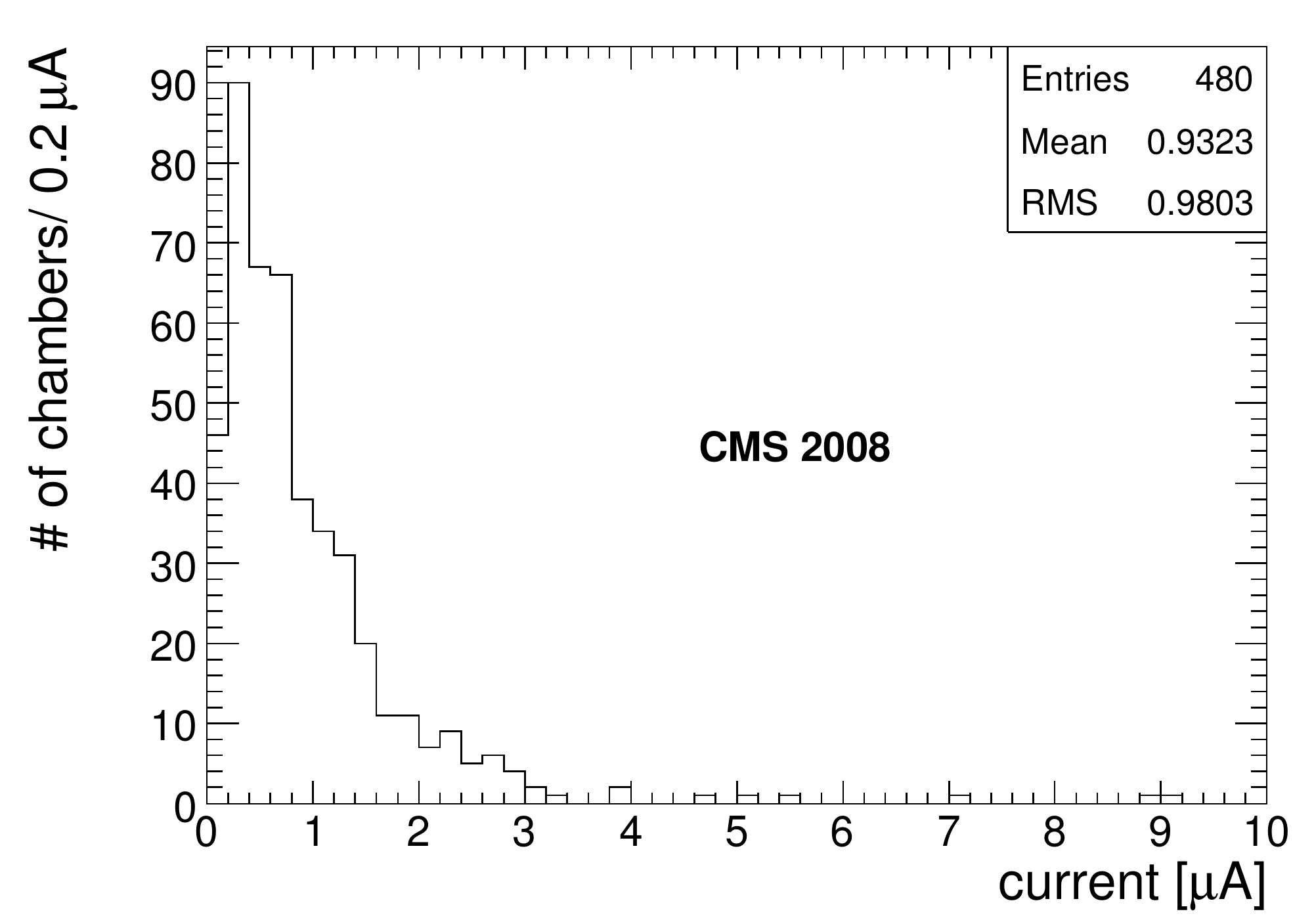} &
    \includegraphics[width=0.45\textwidth]{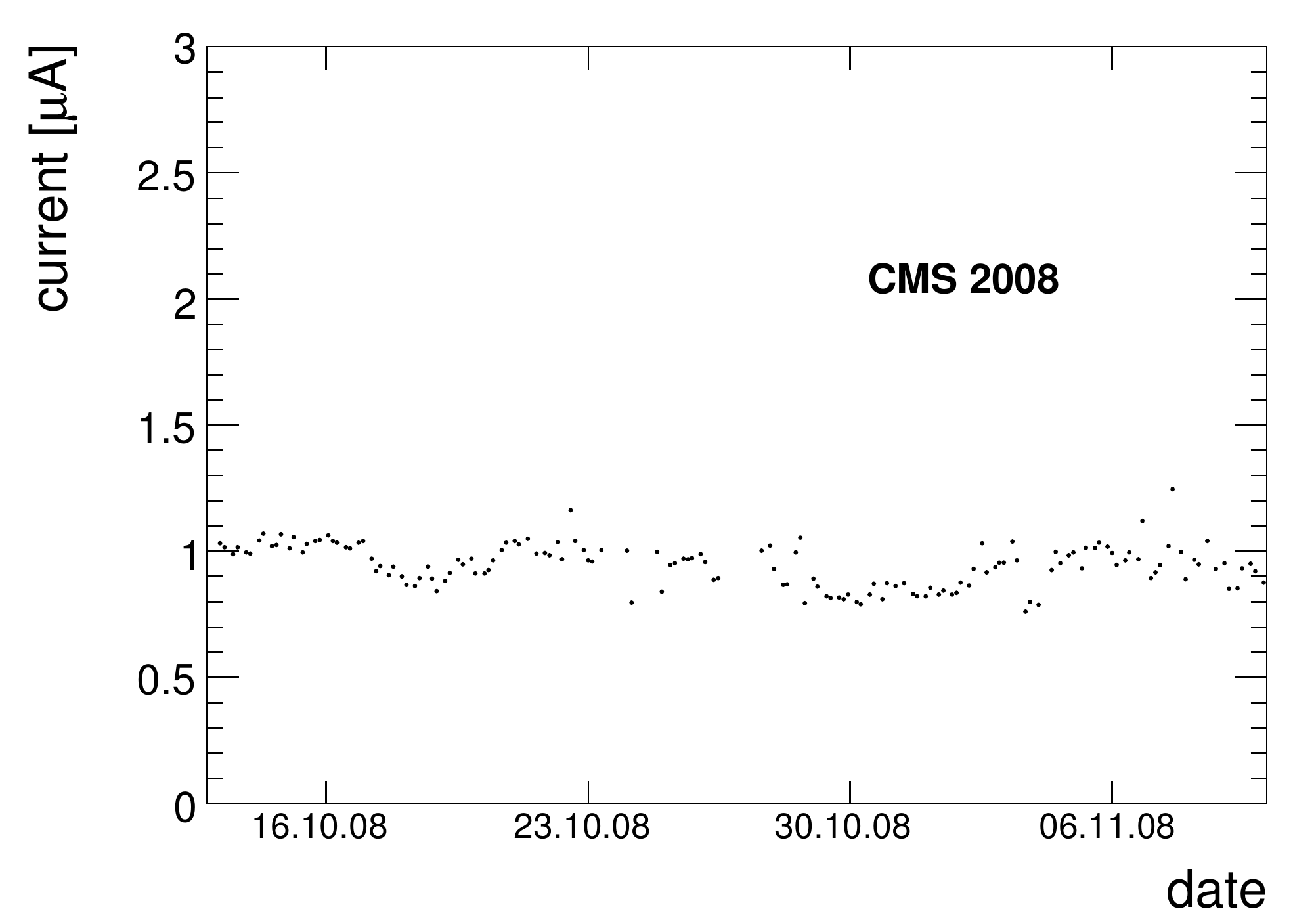} \\
\end{tabular}
  \caption{Distribution of the current drawn by the RPCs.
  Left plot shows the average over one full day of data taking at the end
      of the CRAFT period. The plot
has no overflow. Right plot: average current as a function of time. Each point
      is the average current over 4 hours and over all barrel
      stations.  The operating voltage is 9.2~kV.}
\label{fig:darkCurrent}
\end{center}
\end{figure}

The currents drawn by the RPCs, {\em I}, are strongly affected by
variations of the environmental conditions.  A sudden current increase
could also indicate a malfunctioning of the detector.  Therefore, a
careful and continuous monitoring of the currents is performed.
Each RPC station is powered by a separate high voltage supply.  The
current drawn in each RPC station is read out and stored
through the DCS.  Figure~\ref{fig:darkCurrent} shows in the left plot the {\em I}
distribution at an operating voltage of 9.2~kV, for the 480 barrel
stations, at the end of the CRAFT period. Only for 11 stations the
value of {\em I} is larger than 3~$\mu$A. During the CRAFT exercise, ten
units out of 2040 were disconnected because of problems with the high
voltage connections, leaving the corresponding rolls working in single
gap mode.  Figure~\ref{fig:darkCurrent} right plot shows that the average
current drawn by the stations was around 1~$\mu$A and stable in time.
No correlation is observed between the mean current values and the temperature
variations, in the operating range.

\subsection {Counting rates}

The RPC noise rate is carefully monitored, since abnormal values in
some parts of the detector can result in high fake muon trigger rates.
At the beginning of each run, dedicated calibration data were taken
and analysed to identify and mask noisy readout channels with rates
above 100~Hz/cm$^{2}$. This value was chosen because it
corresponds to the limit that the trigger system can sustain.  The
total number of masked and dead channels was stable during the CRAFT
period, at about 1\,\% of the total number of channels in both
cases.

\begin{figure}[hbtp]
  \begin{center}
    \includegraphics[width=0.55\textwidth, angle=90]{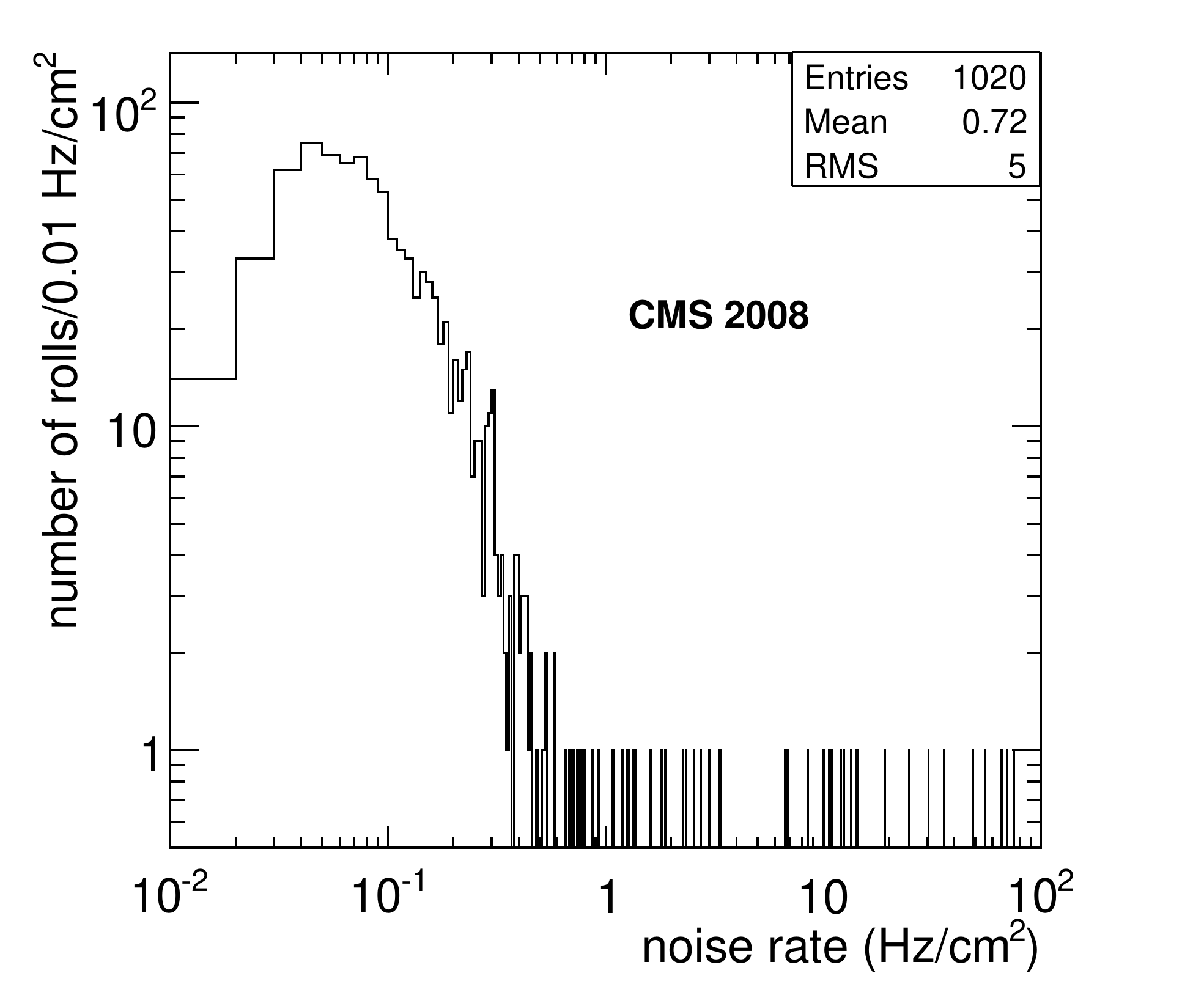}
    \hspace{1cm}
    \caption{Example distribution of the number of rolls as a function
      of the average noise in the roll, for a given data taking run.
      Channels exceeding 100 Hz/cm$^{2}$ are masked and do not
      contribute to the average noise. The operating voltage is
      9.2~kV.}
    \label{fig:noiseByRoll}
  \end{center}
\end{figure}

The average noise per second and per $\rm {cm^2}$ is computed for each
roll by adding the contributions from all non-masked channels.  The
distribution of the number of rolls as a function of the average noise
is shown in Fig.~\ref{fig:noiseByRoll}, for one specific
run.  Only 3\,\% of the rolls have an average noise rate greater than
1~Hz/cm$^{2}$.

Although the average noise rate is very low, a problem
related to events with correlated noise on several layers has
been detected few times per day affecting about $10^{-3}$ of the total data taking time
during the CRAFT exercise. A special
effort has been made to understand the sensitivity of the RPCs to
external noise sources.  The detector grounding has been improved
where possible, and further studies are in progress.  Preliminary
analyses have demonstrated that the fake trigger rate is
reduced by two orders of magnitude when the LHC trigger algorithm is
used instead of the dedicated cosmic ray trigger employed during the CRAFT
exercise.  The LHC trigger algorithm requires more stringent
constraints on the incoming muon direction and is less sensitive to
the detector noise.


\subsection{Gas monitoring}

The mixture composition and its quality are monitored by two
independent devices.  The Gas Quality Monitor
\cite{Abbrescia:2008zzb,Abbrescia:2006hk} performs chemical analyses
(mainly gas chromatography, pH and fluoride monitoring).  The Gas Gain
Monitor~\cite{Benussi:2008fp}, is based on the
monitoring of the performance of three sets of 50 $\times$ 50 cm$^{2}$
single gap RPCs, each supplied with a different gas mixture (fresh
gas, and gas from the recirculating system, before and after the
purifiers).

During the CRAFT exercise, the gas system was operated continuously
with a fraction of 8\,\% of fresh mix.  The lifetime of the purifiers
before regeneration was about 36 hours between regenerations of first
stage purifier, to remove water from gas mixture humidity, and one
week between regenerations of second stage purifier, to remove air
contamination.  The full system was running smoothly during the
operation. The downtime due to gas system stops was less than 1\,\%.

\section{Synchronization of RPC data}\label{sec:synch}

Muon signals detected by the RPC chambers and transmitted by the
readout electronics must arrive in dedicated electronic boards at a
specified time, in order to correctly contribute to the trigger
decisions.  In addition, RPC data sent to the CMS Data Acquisition
system (DAQ) and RPC trigger information must be associated to the
right LHC bunch crossing (BX: 25~ns time interval).

The different steps of data transmission in the RPC system are
described in detail in Ref.~\cite{RPCsynchro}.  After pulse discrimination
in the front end boards, the data are transmitted asynchronously to
the Link Boards situated on the CMS detector, where zero-suppression
is performed. The data are then sent through optical fibres to the CMS
counting room, in the underground cavern, where RPC muon trigger objects
are identified, as part of the CMS Level-1 trigger system, and sent to
the Global Muon Trigger. In parallel, the data are sent to the CMS DAQ
system.  To ensure integration with CMS, the RPC readout and trigger
systems work in synchronous mode using the CMS Timing, Trigger and
Control (TTC) system~\cite{L1TCS}.


The TTC signals (including clock and reference bunch crossing
identification) are distributed to the Link Boards, where the signals
from the detector are synchronized with the LHC clock. The data are
assigned to the proper bunch crossing according to a time window
determined by the length of the TTC fibres, the muon time of flight,
and the signal propagation and transmission times along the strips,
cables and electronics.

The RPC readout~\cite{RPCreadout} is designed to send data from up to
8 consecutive bunch crossings (BXs) to the DAQ system: the trigger BX plus 3
pre- and 4 post-trigger BXs.  During the CRAFT exercise, only 7 BXs
were transmitted to the DAQ ($\pm$3 BXs around the triggered event).

The timing setups of the system are different for the LHC beam mode
and for the cosmic ray runs.  In beam mode the synchronization is driven
by beam collisions, which are fixed in phase with respect to the LHC
clock, while in the cosmic mode this phase is arbitrary: cosmic rays
arrive randomly with respect to the LHC clock.  The synchronizations
of the upper and lower parts of the detector differ significantly in
the two modes, due to the
asymmetric phi distribution of the cosmic rays.

\begin{figure}[hbtp]
  \begin{center}
    \includegraphics[width=0.5\textwidth, angle=90]{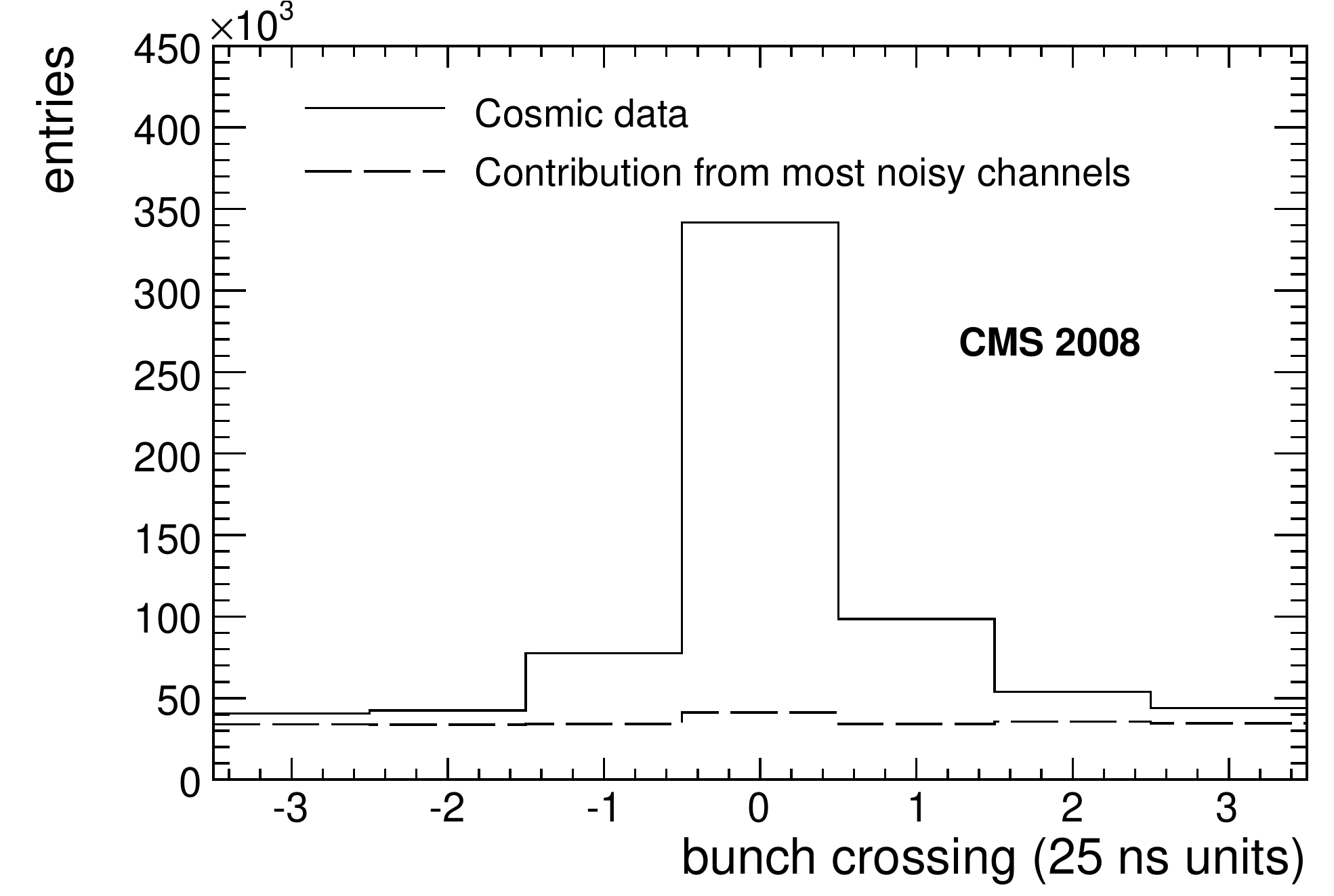}
    \hspace{1cm}
    \caption{Typical time distribution (in bunch crossing units of
      25~ns) of the data coming from individual RPCs with
      respect to the RPC global trigger time. The dashed line shows
      the background caused by the 10 over 1020 Link Boards connected to the most
      noisy channels.}
    \label{fig:delaySummary}
  \end{center}
\end{figure}

In order to optimize the data synchronization, initial delays were
estimated, based on the length of cables and fibres.  The
bottom-central part of the detector was chosen as a reference, and
delays of data from other parts of the detector were fine-tuned to
optimize the data alignment in time.  The same timing was kept for the
full CRAFT period.  As an illustration of the RPC synchronization
during the CRAFT exercise, Fig.~\ref{fig:delaySummary} presents a
typical example of the time distribution of the data coming from
individual RPCs. The time is counted with respect to the
bunch crossing assigned by the RPC Level-1 global trigger logic, based
on the coincidence of at least 3 chambers along a muon trajectory.
The peak in the central bin corresponds to data which are synchronous
with the RPC Level-1 trigger. The spread to the two neighbouring bins is
caused mainly by the arbitrary choice of the synchronization phase,
relevant only for cosmic ray runs. The spread over the other bins is due
to the non-perfect synchronization of contributing triggers, and to a
background caused by Link Boards connected to the most noisy
electronic channels.

In view of the good quality of the present data, much better
synchronization is expected for LHC runs.

\section{Detector performance}\label{sec:performances}

The performance of the RPC system has been studied by making use of
the local DT hit reconstruction~\cite{dt}.  The three innermost DT
layers are able to locally reconstruct three dimensional muon track
segments with up to eight hits in the $r\phi$ plane and up to four
hits in the $rz$ plane.  The outermost DT layer provides a segment in
the $r\phi$ plane.


The extrapolation of a DT track segment onto an RPC plane allows the
study of the local RPC performance, by searching for RPC channels over
threshold in a small region around the impact point.  Previous studies
\cite{MTCC} have shown that a range of $\pm$ 2 strips between the
extrapolated impact point and the closest firing strip is adequate.

In addition, the CRAFT exercise led to the identification of 15
swapped readout cables out of a total of 4720, and a few cases of RPC
to DT misalignments at the level of more than 1~cm.

In the following sub-sections the results on cluster size, position
resolution and detection efficiency are reported for the whole barrel
system.  The RPC cluster size and the position resolution are studied
separately for layers with different strip pitches.  No special
selection is applied to the DT track segments used for the
extrapolation, except for the requirement that no other segment is
reconstructed in the DT station. This condition is imposed to reject
multiple muon events and other possible ambiguous topologies.
Moreover, only pointing segments in the $r\phi$ plane have been used
in these analyses. A pointing track segment in the $r\phi$ plane is defined
as a segment within an angle of $\pm$ 20 degrees around the normal to
the RPC layer, in the plane perpendicular to the strip direction.  All
the figures have been produced by analysing a run of about 2 million
events, representative of the full CRAFT period.

\subsection{Cluster size}

A cluster is defined as a consecutive set of strips, each of them
collecting an induced charge above the discrimination threshold of
180~fC. The number of strips in the cluster is called the cluster
size.

\begin{figure}[hbtp]
  \begin{center}
    \includegraphics[width=0.55\textwidth, angle=90]{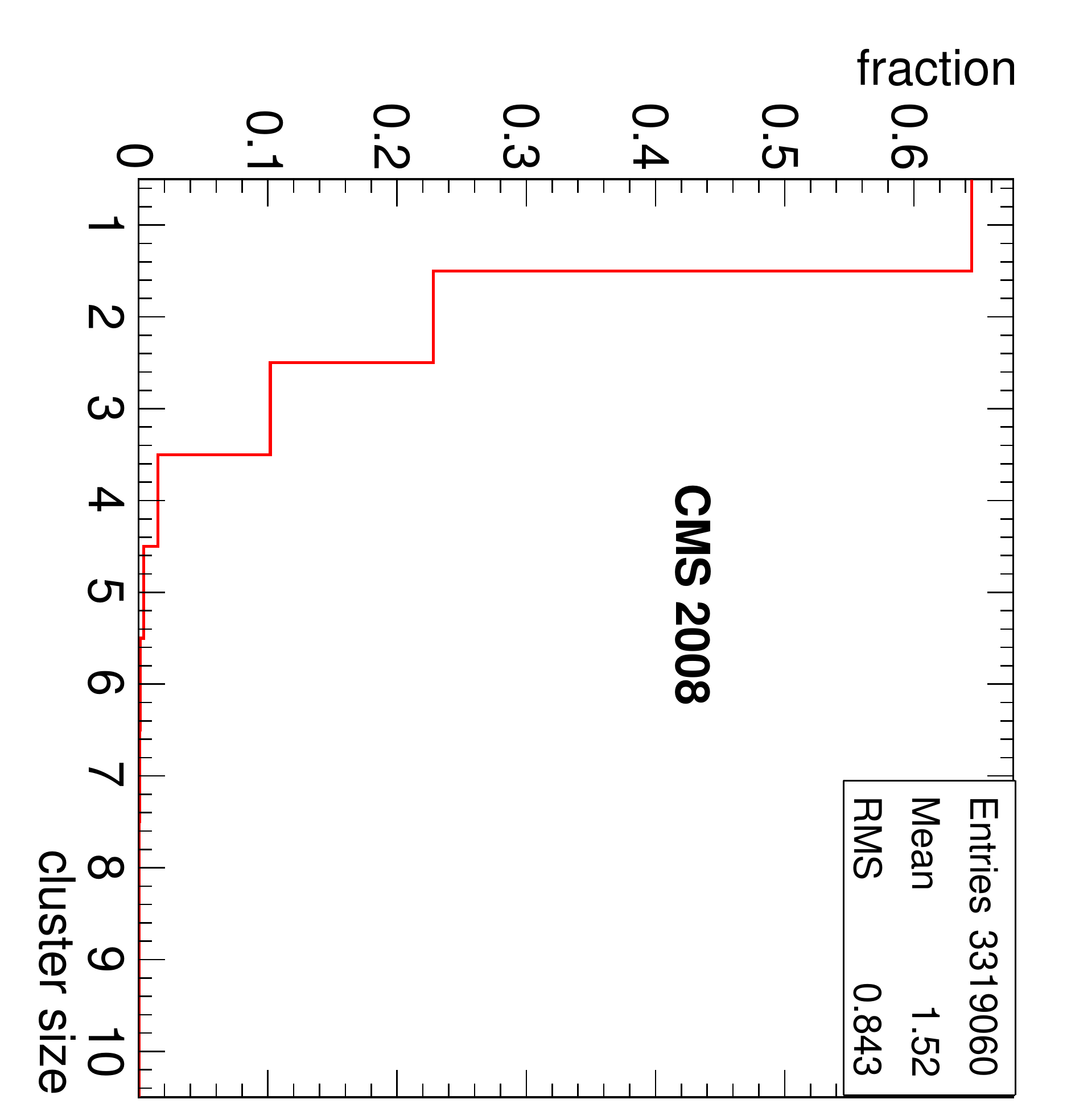}
    \hspace{1cm}
    \caption{Normalized distribution of the cluster size for the RB1$_{\rm in}$
      layer, for pointing track segments.}
    \label{fig:ClusterSizeForBarrel}
  \end{center}
\end{figure}

\begin{figure}[hbtp]
  \begin{center}
    \includegraphics[width=0.55\textwidth, angle=90]{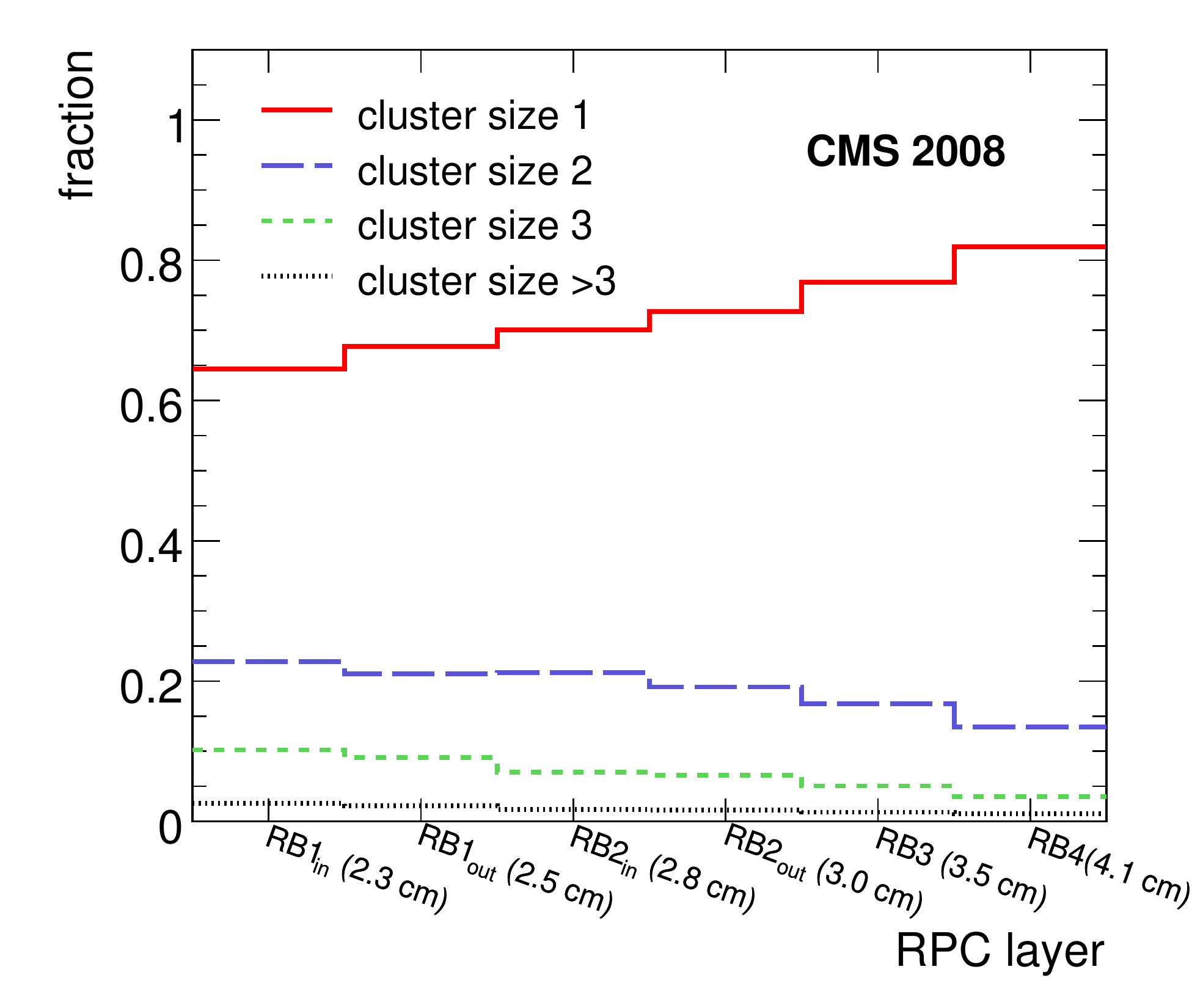}
    \hspace{1cm}
    \caption{Relative population of reconstructed clusters with size equal to 1,
      2, 3 and more than 3 strips, for each RPC layer.}
    \label{fig:clsByLayer}
  \end{center}
\end{figure}

The cluster size depends on the RPC strip pitch, on the impact point
position with respect to the strip, and on the track crossing angle.
Figure~\ref{fig:ClusterSizeForBarrel} shows the cluster size
distribution for the RB1$_{\rm in}$ layers, for pointing muons.  The
different strip pitches and the average cluster sizes are given for
all layers in Table~\ref{tab:tabclsByLayer}. The fractions of events
having cluster sizes corresponding to 1, 2, 3 or more than 3 strips
are shown for different layers in Fig.~\ref{fig:clsByLayer}.

\begin{table}[htbH]
\begin{center}
  \caption{Strip pitch and average cluster size, counted as the number
    of strips, for pointing track segments, with operating voltage of
    9.2~kV.\label{tab:tabclsByLayer}}
\begin{tabular}{|l|c|c|}
\hline
RPC layer & strip pitch (cm) & average cluster size  \\
\hline\hline
RB1$_{\rm in}$ & 2.3 & 1.52  \\
RB1$_{\rm out}$ & 2.5 & 1.46  \\
RB2$_{\rm in}$ & 2.8 & 1.41  \\
RB2$_{\rm out}$ & 3.0 & 1.38  \\
RB3 & 3.5 & 1.31  \\
RB4 & 4.1 & 1.25  \\
\hline
\end{tabular}
\end{center}
\end{table}

\begin{figure}[hbt]
  \begin{center}
    \includegraphics[width=0.6\textwidth, angle=90]{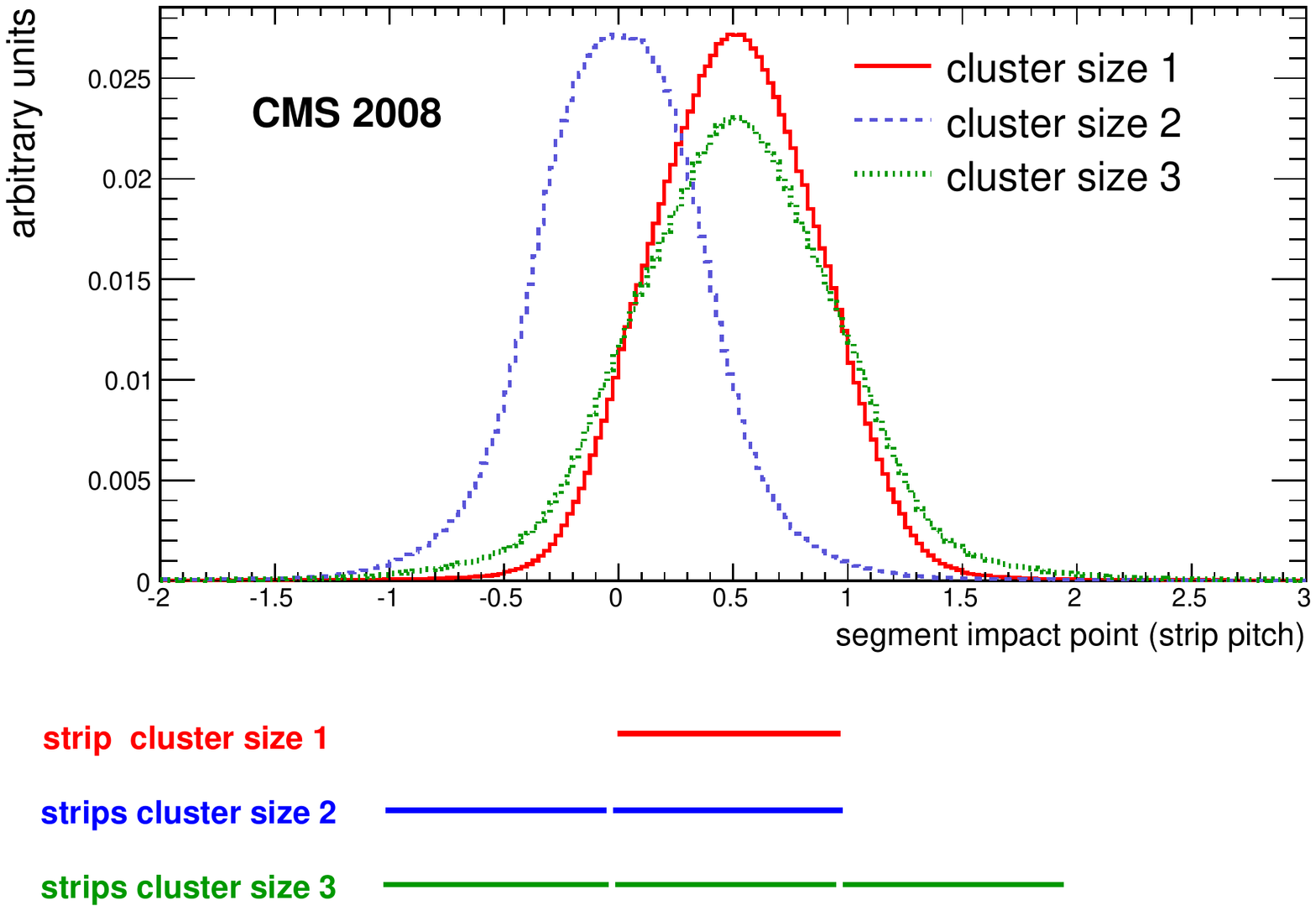}
    \hspace{1cm}
    \caption{Distribution of the track impact
      point for the RB1$_{\rm in}$ layer in a local
      reference frame, for events with cluster size of 1, 2 and 3, in
      units of strip pitch. The local frame, described by the drawing under the plot, is defined in such a way
      that the fired strip is between 0 and 1 for events with cluster
      size of 1, the two fired strips are between $-$1 and 1 for events
      with cluster size of 2, and the three fired strips are between
      $-$1 and 2 for events with cluster size of 3. The three
      distributions are normalized to the same unit area.}
    \label{fig:clsVsImpactPoint}
  \end{center}
\end{figure}

The number of firing strips for a crossing muon is a function of the
local impact point position with respect to the RPC strip.  The
distribution in Fig.~\ref{fig:clsVsImpactPoint} shows the deviation,
$\Delta$, measured in strip pitch units, between the impact point and
the cluster centre for the RB1$_{\rm in}$ layer, in a local reference
frame.  In this local frame the fired strip is between 0 and 1 for
events with cluster size of 1, the two fired strips are between $-$1 and
1 for events with cluster size of 2, and the three fired strips are
between $-$1 and 2 for events with cluster size of 3.  The three
distributions, normalized to the same area, are overlayed.  Events
with cluster size of 1 are more frequent for tracks crossing the RPC
layer close to the middle of a strip, while the fraction of events
with a cluster size of 2 increases for muons close to the edge between
two adjacent strips.

\begin{figure}[hbtp]
  \begin{center}
    \includegraphics[width=0.55\textwidth, angle=90]{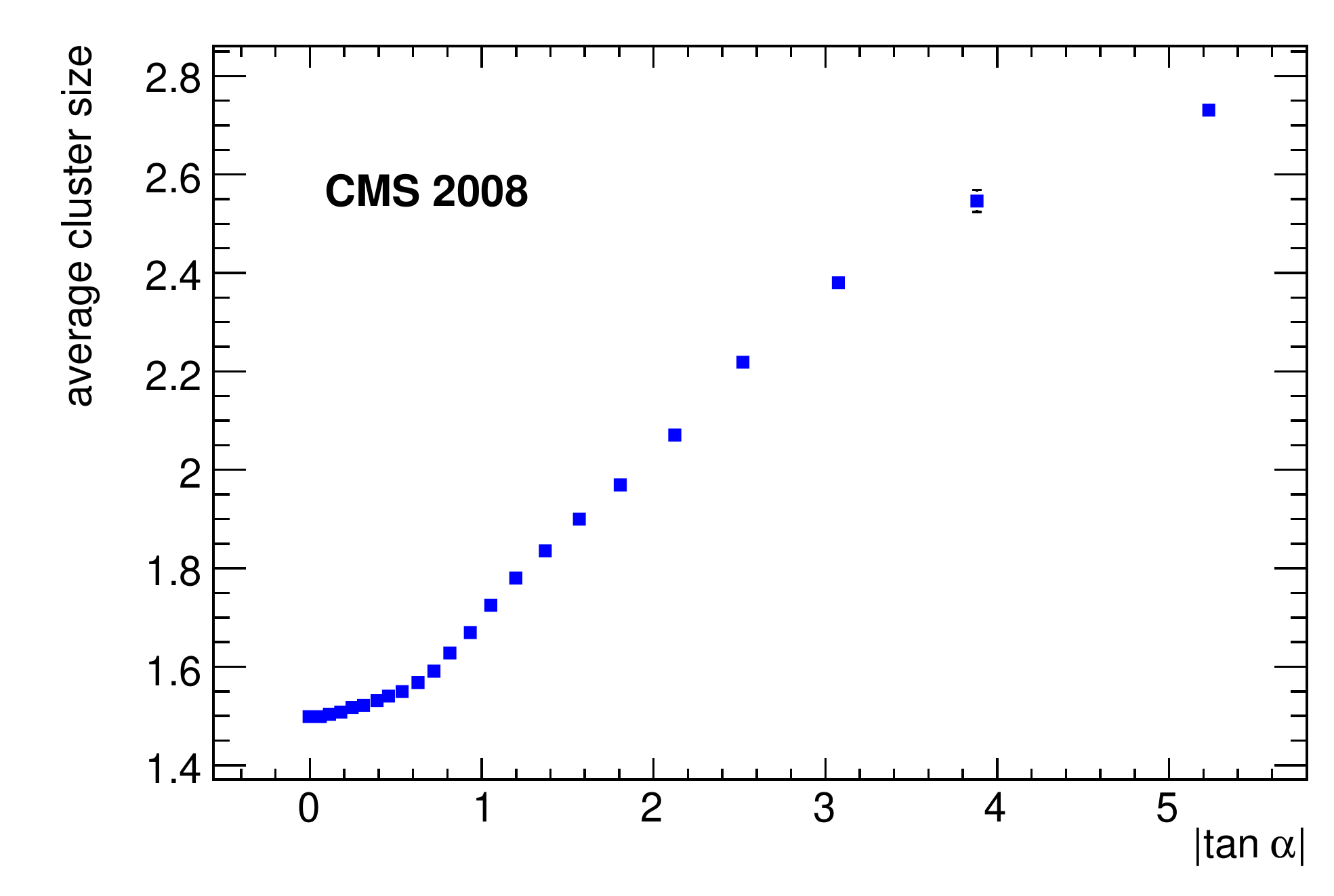}
    \hspace{1cm}
    \caption{Average cluster size as a function of the absolute value
      of the tangent of the angle $\alpha$, defined as the angle
      between the track direction and the normal to the RPC surface in
      the plane perpendicular to the strip direction, for the
      RB1$_{\rm in}$ layer and with an operating voltage of 9.2~kV.}
    \label{fig:clsVsAngle}
  \end{center}
\end{figure}

In addition, the cluster size depends on the muon incident angle with
respect to the RPC surface, in the plane orthogonal to the strip
direction. This is visible in Fig.~\ref{fig:clsVsAngle}, which
presents the dependence of the average cluster size on the absolute
value of the tangent of the angle $\alpha$, defined as the angle
between the muon direction and the normal to the RPC surface in the
plane perpendicular to the strip direction, for the RB1$_{\rm in}$ layer.
This dependence is due both to the increased path length in the
crossed gas gap and to the RPC double gap structure where two
independent avalanches develop in the two gas gaps, 6~mm apart, which
may induce signals on different strips, increasing the cluster size.

It should be noted that, because of the bending of muon trajectories
due to the magnetic field, low momentum muons coming
from the interaction point, during LHC runs, may cross the RPC detectors with relatively
large angles.  Cosmic muons are very useful to study these topologies
since they cross the detector at all possible angles, depending on the sector.

\subsection{Position resolution}

The centre of a cluster provides a measurement of the position of the
muon track crossing point, in the $r\phi$ plane.  In order to
study the RPC position resolution, the cluster centre is compared to
the extrapolated point of the DT segment onto the RPC layer.  The
distribution of the off-set between the position of the track
crossing point onto the surface separating the two gaps and the
cluster centre, named residual, is computed for every roll, providing
an estimate of the relative DT and RPC alignment.

\begin{figure}[hbtp]
  \begin{center}
    \includegraphics[width=0.55\textwidth, angle=90]{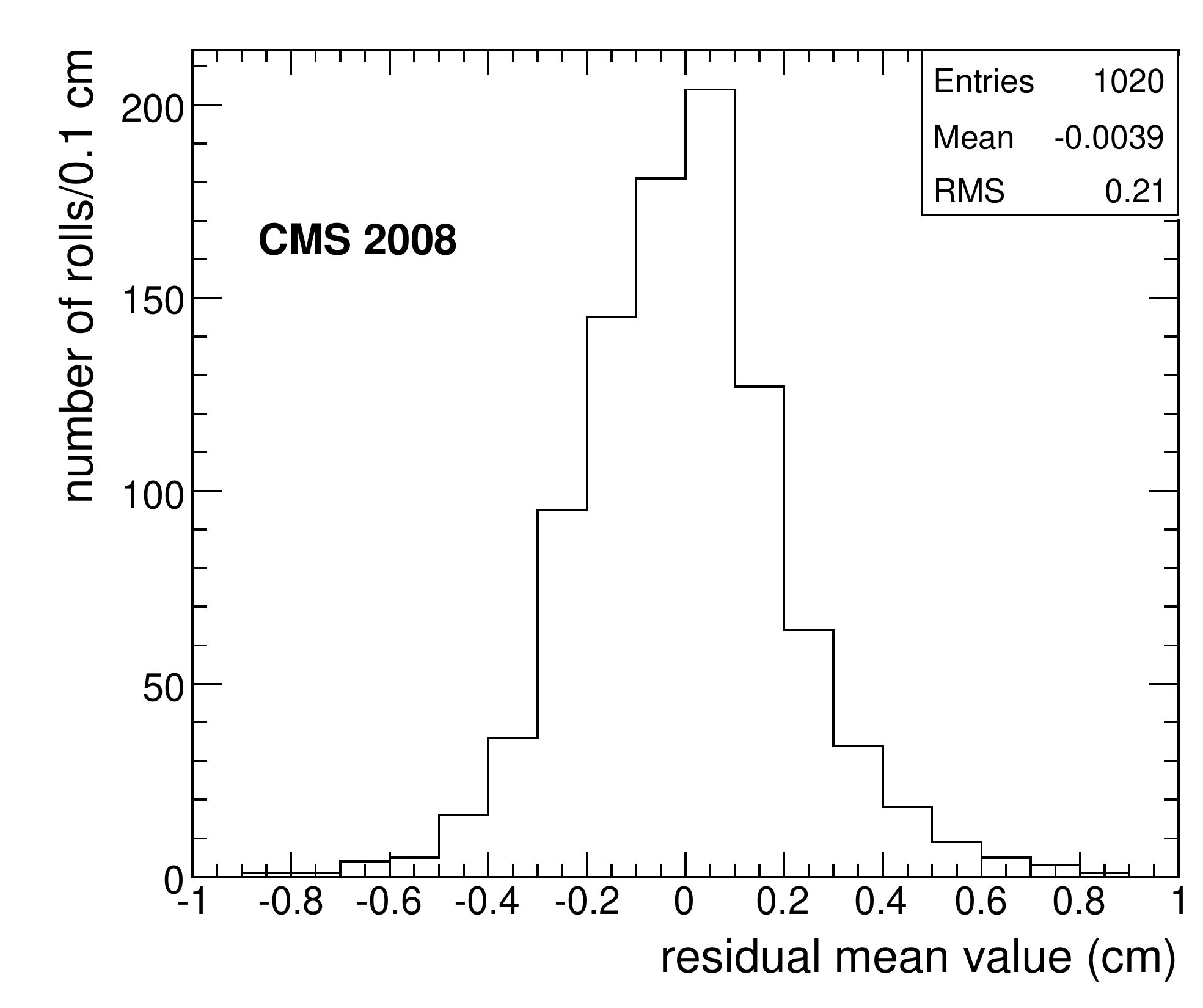}
    \hspace{1cm}
    \caption{Normalized distribution of the mean value of the residuals in~cm, for all
      the RPC rolls. Two entries are out of the range.}
    \label{fig:offset}
  \end{center}
\end{figure}

Figure~\ref{fig:offset} presents the distribution of the mean value of
the residuals for all barrel rolls.  The relative alignment is of a
few millimetres, but two cases of a few centimetres have also been
observed (not visible on the plot scale).  The mean values extracted
from the residual distributions have been used to correct the data
offline, in order to improve the position resolution.

\begin{figure}[hbtp]
  \begin{center}
    \includegraphics[width=0.55\textwidth, angle=90]{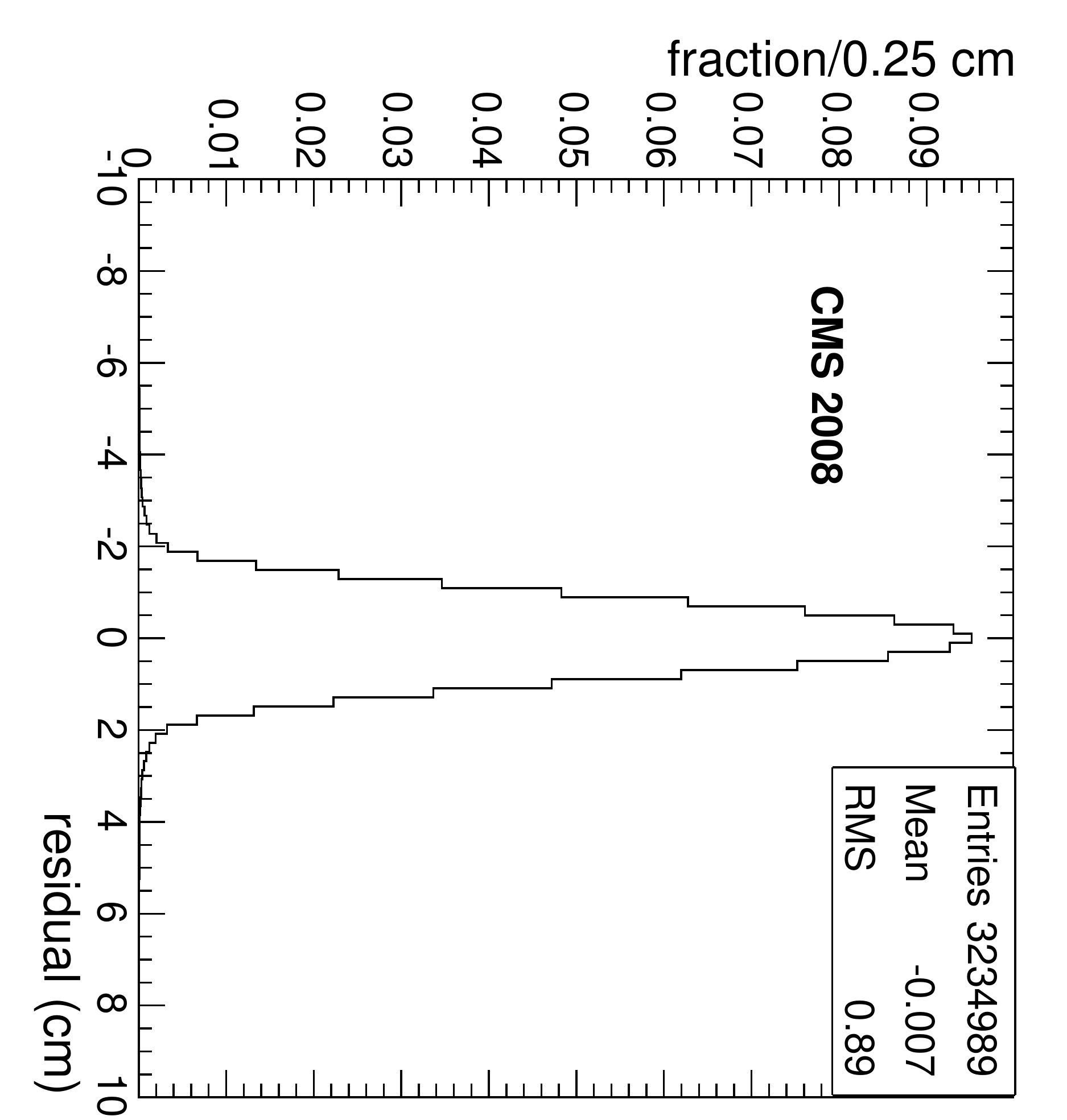}
    \hspace{1cm}
    \caption{Normalized distribution of the residuals in~cm, for the RB1$_{\rm in}$ layer
      (events with cluster size larger than three are not included).}
    \label{fig:residuals}
  \end{center}
\end{figure}

Figure~\ref{fig:residuals} shows the residual distribution in~cm for the
RB1$_{\rm in}$ layer, after having applied the alignment corrections, for
events with cluster size lower than four.

\begin{figure}[hbtp]
  \begin{center}
    \includegraphics[width=0.55\textwidth, angle=90]{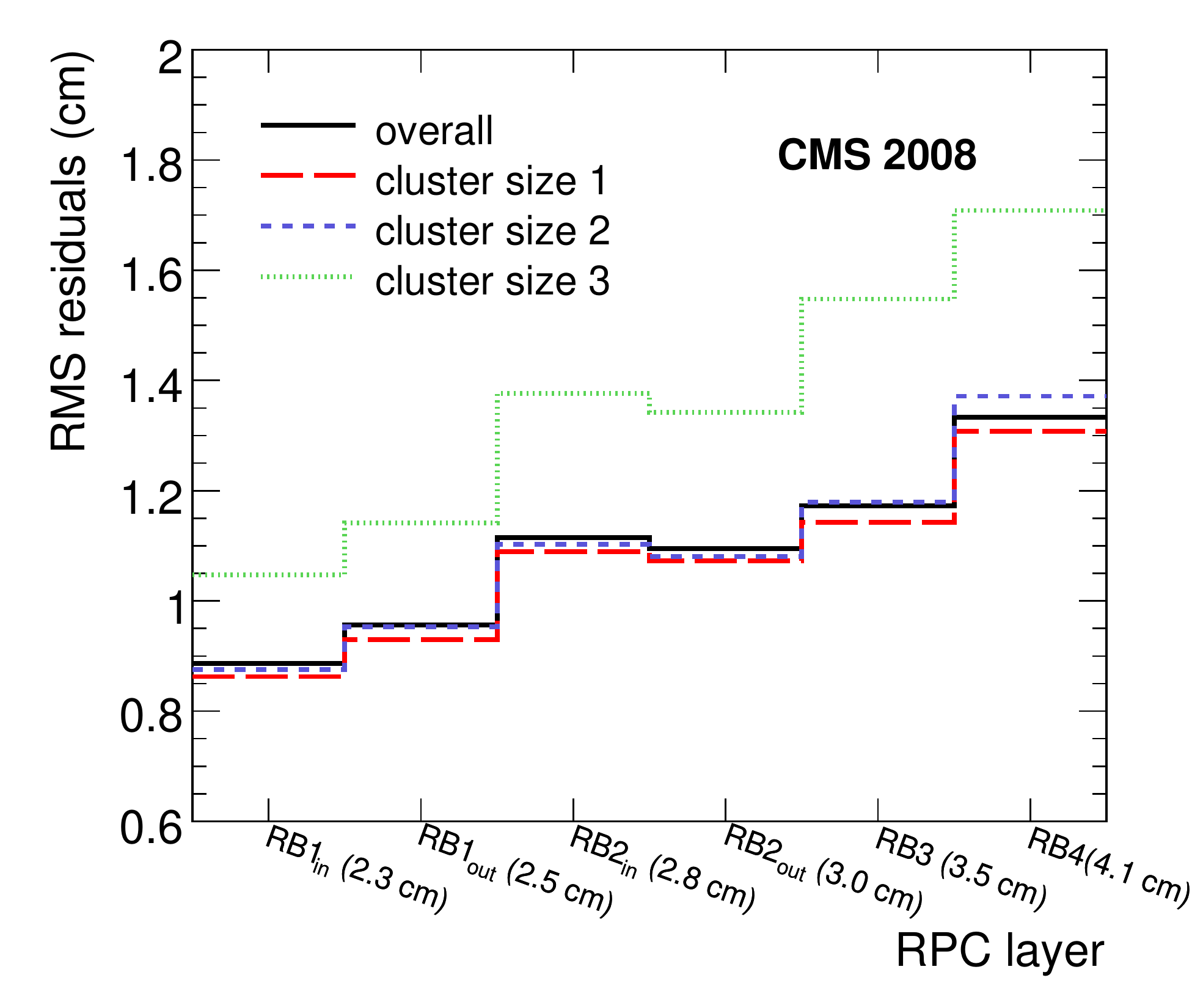}
    \hspace{1cm}
    \caption{RMS of the distributions of the residuals, for the
      different layers (corresponding to different strip pitches) and
      for different cluster sizes.}
    \label{fig:RMSByLayer}
  \end{center}
\end{figure}

The analysis of the RPC residuals has been performed for all layers,
corresponding to different strip pitches as reported in Table
\ref{tab:tabclsByLayer}, and for events with different cluster sizes.
The RMS values of the residual distribution are
presented in Fig.~\ref{fig:RMSByLayer}, as a function of the layer, for different cluster sizes.
The trend reflects the variation of the strip pitch and the geometry
of the different RPC layers.

\subsection{Detection efficiency}

The most important parameter defining the RPC performance is the
detection efficiency.  The DT segment extrapolation, presented at the
beginning of Section~\ref{sec:performances}, is used to estimate the
RPC efficiency, for a full roll.

\begin{figure}[hbt]
  \begin{center}
    \includegraphics[width=1.\textwidth]{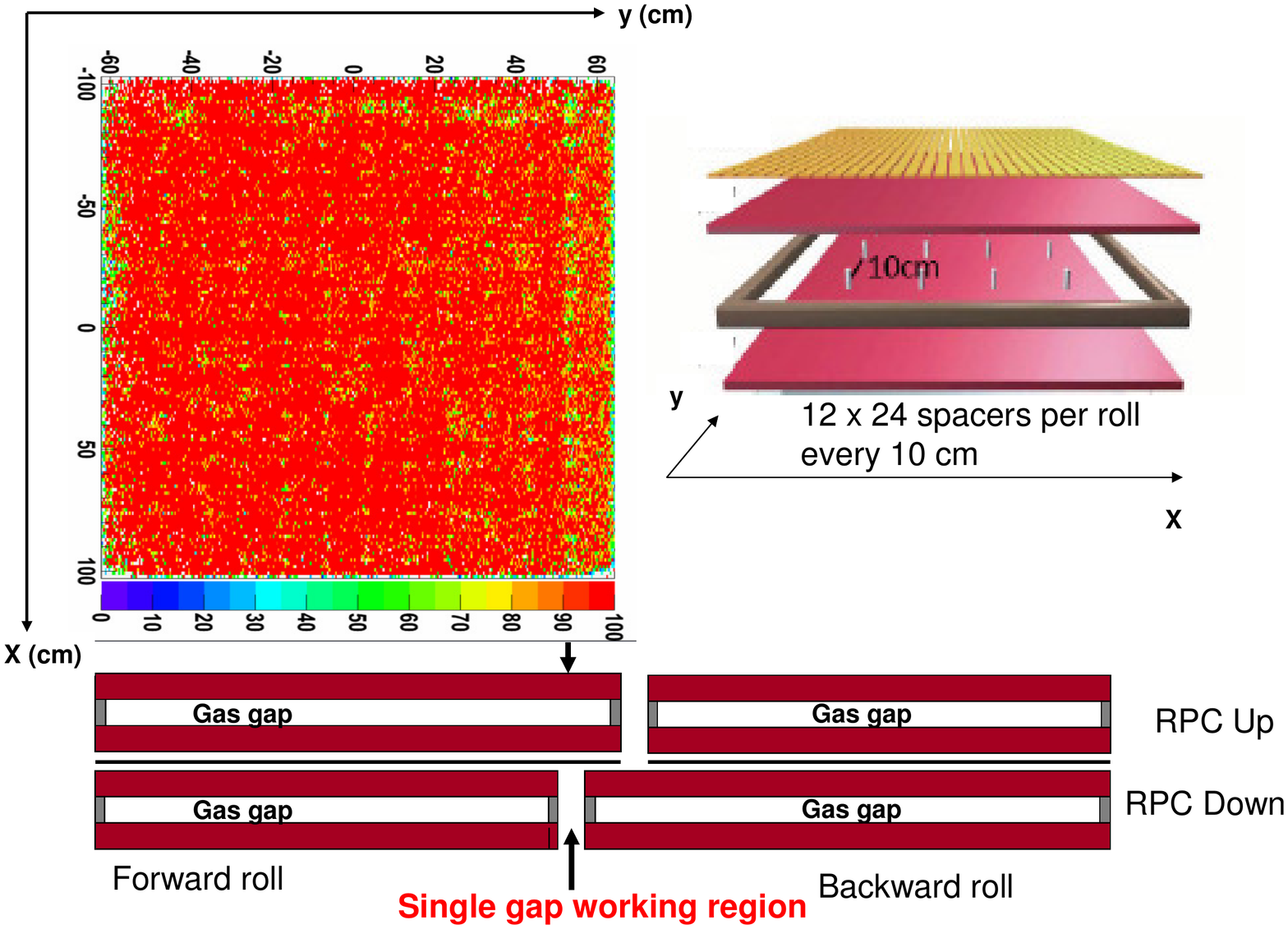}
    \hspace{1cm}
    \caption{Efficiency of an RB1$_{\rm in}$ roll as a function of the
      $x$ and $y$ coordinates (as defined in the figure) of the
      extrapolated track impact point.  The lower efficiency spots are
      due to the dead regions induced by spacers on a 10 $\times$ 10
      cm$^{2}$ grid.  An efficiency reduction is also visible for the
      $y$ coordinate at about 55 cm, where only a single gap is
      present.}
    \label{fig:spacersXY}
  \end{center}
\end{figure}

For each DT track segment extrapolated onto the RPC surface, a roll is
considered efficient if at least one RPC strip is over threshold
within $\pm$~2 strips from the extrapolated
impact point.  Figure~\ref{fig:spacersXY} shows the efficiency vs.\ the
local impact point on the RPC surface.  The pattern due to the spacers
placed on a 10 $\times$ 10 cm$^{2}$ grid is clearly visible.  Moreover,
a reduction of the overall efficiency is visible in the $y$ coordinate
(as defined in the figure) at about 55~{\rm cm}, where only one gap is
present.

\begin{figure}[hbtp]
  \begin{center}
    \includegraphics[width=.55\textwidth, angle=90]{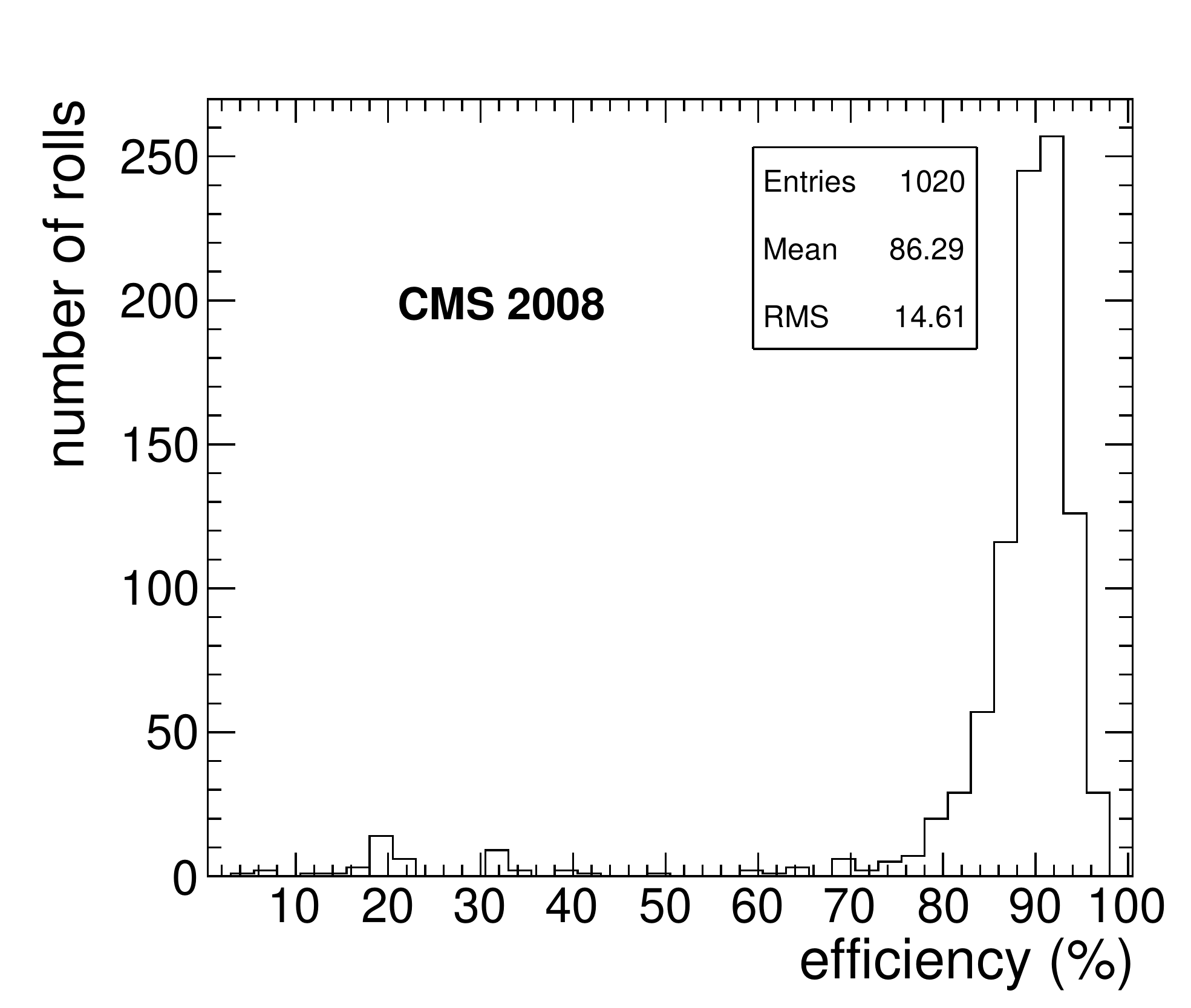}
    \hspace{1cm}
    \caption{Efficiency distribution for barrel rolls, at an operating
      voltage of 9.2~kV.  The tail at lower efficiency values is due
      to now known swapped cables, chambers working in single gap mode
      and chambers with synchronization problems.}
    \label{fig:eff9200}
  \end{center}
\end{figure}

The roll efficiency distribution is shown in Fig.~\ref{fig:eff9200}
for a typical run, taken at the operating voltage of 9.2~kV.  The peak
of the distribution is about 90\,$\%$, while the tail at lower
efficiency values is due to now known swapped cables, chambers working
in single gap mode and chambers with synchronization problems.

After the end of the CRAFT exercise many improvements have been obtained and
the operating conditions have been optimized. In 2009 the efficiency distribution
was around 95\,$\%$ at 9.4~kV.

\section{Conclusions}\label{sec:conclusions}

Data collected during the CRAFT 2008 period have been very useful for
detector commission\-ing. About 98\,$\%$ of the electronic channels were operational during the
exercise.
 Hardware problems such as swapped cables and
electronic failures have been identified and fixed. The relative
position of each RPC station with respect to the DT chambers has been
measured in the $r\phi$ direction, and the position resolution of the
RPC system has been determined.  Software tools to monitor the system
performance have been developed and tuned during the running
period. The performance of the barrel system has been measured
from data. The results show a good stability of the detector and an
efficiency around 90\,$\%$ at the operating voltage of 9.2~kV.  Many
improvements have been made after the CRAFT 2008
exercise. The temperature of the input cooling water has been reduced by
one degree. Hardware failures have been fixed, the working voltage and
the electronic threshold have been optimized, and the synchronization of
the RPC barrel signals has been improved.  Although more systematic
studies are still in progress, the RPC barrel system is ready for LHC
beam collisions.

\section*{Acknowledgements}

We thank the technical and administrative staff at CERN and other CMS
Institutes, and acknowledge support from: FMSR (Austria); FNRS and FWO
(Belgium); CNPq, CAPES, FAPERJ, and FAPESP (Brazil); MES (Bulgaria);
CERN; CAS, MoST, and NSFC (China); COLCIENCIAS (Colombia); MSES
(Croatia); RPF (Cyprus); Academy of Sciences and NICPB (Estonia);
Academy of Finland, ME, and HIP (Finland); CEA and CNRS/IN2P3
(France); BMBF, DFG, and HGF (Germany); GSRT (Greece); OTKA and NKTH
(Hungary); DAE and DST (India); IPM (Iran); SFI (Ireland); INFN
(Italy); NRF (Korea); LAS (Lithuania); CINVESTAV, CONACYT, SEP, and
UASLP-FAI (Mexico); PAEC (Pakistan); SCSR (Poland); FCT (Portugal);
JINR (Armenia, Belarus, Georgia, Ukraine, Uzbekistan); MST and MAE
(Russia); MSTDS (Serbia); MICINN and CPAN (Spain); Swiss Funding
Agencies (Switzerland); NSC (Taipei); TUBITAK and TAEK (Turkey); STFC
(United Kingdom); DOE and NSF (USA). Individuals have received support
from the Marie-Curie IEF program (European Union); the Leventis
Foundation; the A. P. Sloan Foundation; and the Alexander von Humboldt
Foundation.

\bibliography{auto_generated}   
\clearpage
\appendix

\cleardoublepage\appendix\section{The CMS Collaboration \label{app:collab}}\begin{sloppypar}\hyphenpenalty=500\textbf{Yerevan Physics Institute,  Yerevan,  Armenia}\\*[0pt]
S.~Chatrchyan, V.~Khachatryan, A.M.~Sirunyan
\vskip\cmsinstskip
\textbf{Institut f\"{u}r Hochenergiephysik der OeAW,  Wien,  Austria}\\*[0pt]
W.~Adam, B.~Arnold, H.~Bergauer, T.~Bergauer, M.~Dragicevic, M.~Eichberger, J.~Er\"{o}
% ...
\vskip\cmsinstskip
% ...

\end{sloppypar}
\end{document}